\documentclass[12pt]{article}
\usepackage{newinutile-pre}
\usepackage{amsmath,amssymb}
\usepackage{epsfig,graphics,color,calc,graphicx}
\usepackage{cite}
\usepackage{mathrsfs}

\newlength{\pecettawidth}
\begin{document}
\title{Free to move or trapped in your group: 
Mathematical modeling of information
overload and coordination in crowded populations}

\author{A.\ Ciallella}
\email{alessandro.ciallella@uniroma1.it}
\affiliation{Dipartimento di Scienze di Base e Applicate per l'Ingegneria, 
             Sapienza Universit\`a di Roma, 
             via A.\ Scarpa 16, I--00161, Roma, Italy.}

\author{Emilio N.M.\ Cirillo}
\email{emilio.cirillo@uniroma1.it}
\affiliation{Dipartimento di Scienze di Base e Applicate per l'Ingegneria, 
             Sapienza Universit\`a di Roma, 
             via A.\ Scarpa 16, I--00161, Roma, Italy.}

\author{Petru L.\ Cur\c seu}
\email{petrucurseu@psychology.ro}
\affiliation{Department of Psychology Babes--Bolyai University,
Republicii 37, Cluj--Napoca, 400015 Cluj, Romania.
\\
Department of Organization, Open University of The Netherlands, Heerlen, The Netherlands.
}

\author{Adrian Muntean}
\email{adrian.muntean@kau.se}

\affiliation{Department of Mathematics and Computer Science,
             Karlstad University, Sweden.}

\begin{abstract}
We present\footnote{Preprint of an article submitted for consideration in 
Mathematical Models and Methods in Applied Sciences, 2018, 
World Scientific Publishing Company, 
www.worldscientific.com/worldscinet/m3as.}
modeling strategies that describe the motion and interaction of groups of pedestrians in obscured spaces. We start off with an approach based on balance equations in terms of measures and then we exploit the descriptive power of a probabilistic cellular automaton model.
Based on a variation of the simple symmetric random walk
on the square lattice,
we test the
interplay between population size and an interpersonal attraction parameter for
the evacuation of confined and darkened spaces. We argue that information
overload and coordination costs associated with information processing in small
groups are two key processes that influence the evacuation rate. Our results
show that substantial computational resources are necessary to compensate for
incomplete information -- the more individuals in (information processing)
groups the higher the exit rate for low population size. For simple social
systems,
it is likely
that the individual representations are not redundant and large group sizes
ensure that this non--redundant information is actually available to a
substantial number of individuals.  For complex social systems
information
redundancy makes information evaluation and transfer inefficient and, as such,
group size becomes a drawback rather than a benefit. The effect of group sizes on outgoing fluxes, evacuation times and wall effects are carefully studied with a Monte Carlo framework accounting also for the presence of an internal obstacle.
\end{abstract}



    


\maketitle

\section{Introduction}
\subsection{Towards modeling pedestrian group dynamics}
\par\noindent
Social mechanics is a topic that has attracted the attention of
researchers for more than one hundred years; see e.g. \cite{Haret,Apuntes}.
A large variety of existing crowd dynamics models are able to describe the dynamics
of pedestrians driven by a {\em desired velocity} towards clearly defined exits.
But how can we possibly describe the motion of pedestrians
when the exits are unknown or not clearly defined, or even worse, {\em what if the exits are not visible}?

This paper is inspired by a practical evacuation scenario.
Some of the existing models are geared towards describing
the dynamics of pedestrians with somehow given, prescribed or,
at least, desired velocities or spatial fluxes towards an exit the location of which is,
more or less, known to the pedestrians. Our contribution complements the existing work on crowd evacuation dynamics, see e.g. the papers \cite{ACK2015,Albi,BCGTV2016,MWS2014,WS2014,Xue} and references cited therein.

We focus on modeling basic features that we assume to be influential for the motion of pedestrians in regions with reduced or no visibility. In our simulation studies, we focus on group size (social cluster size), population size, interaction with the walls of the confined space and the presence of an obstacle in the room as factors that might impact on information overload and coordination costs in social systems. In terms of output we focus both on the outgoing flux when the population size is constant and the evacuation time of the whole population in the confined space (under this condition, population size decreases constantly with the evacuation flux).

\subsection{A case study in group psychology. Aim of the paper.}
\label{s:intro}
\par\noindent
In a classic study in Social Psychology \cite{sherif}
the autokinetic effect is used 
to explore the formation of norms in social groups. The autokinetic effect is an optical illusion generated by a stationary light spot projected on a display in a darkened room, spot that appears to be moving \cite{adams}. 
In \cite{sherif} participants are asked 
to estimate the amplitude of the movement of this light spot, first individually and then in dyads or groups of three. His observation was that although the individual estimates were substantially different, when group members could hear each other’s answers, their estimations converged towards consensus on the amplitude of the movement. Moreover, when participants were first asked to estimate the amplitude of the movement in a group and then individually, the ``norm'' 
established in the group biased the individual estimations later on. Finally, it is a matter of convergence in time as the individual estimates did not converge after the first trial, in other words, group members needed time to synchronize their individual estimates 
\cite{sherif}. This is one of the first empirical evidences that in ambiguous situations, people often use others as information resources  to find clarity and come to accurate evaluations of environmental stimuli. 
In a similar fashion, in cases of fire emergency, in rooms filled with smoke, people are confronted with highly ambiguous stimuli and attempt to find clarity by interacting with others present in the same room (situation) \cite{latane}.

The amount of information available to the population present in the room, increases with the population size as each member in the room may get access to a particular information that is otherwise inaccessible to the others in the room. In other words, the more people in the space to be evacuated, the higher the potential pool of information for the accurate evaluation of the situation. There are however limitations the information sharing possibilities in groups of varying sizes. As the experiment in \cite{sherif}, 
the individual estimates needed time to converge as during each exposure, participants slowly adjusted their individual estimation to the estimates of the other groups members, till eventually all estimates converged towards a group norm. These coordination constraints of individual actions in groups or communities have subsequently received substantial attention in the literature 
\cite{mueller}. 
People are more likely to share information with the ones situated in their immediate presence, 
therefore information sharing and information integration attempts 
happen in smaller groups or clusters
\cite{curseu2008,curseu2014,hinsz}. 
As the size of these small groups increases so does the coordination cost associated with information sharing and evaluation. 
The way in which people use each other as information resources in ambiguous situations is therefore constrained by at least two important 
socio--cognitive factors. The first one refers to the information pool available to a population exposed to highly ambiguous situations. As the population size increases, so does the potential information pool available in the system. The second one refers to the coordination constraints, as information sharing happens usually in smaller subgroups. As the size of the subgroup in which the information sharing takes place increases, so do the coordination costs associated with information sharing and information evaluation. An important problem therefore arises on how factors that might impose coordination constraints and generate information overload (the population size, the cluster size, interaction with walls and presence of obstacles) influence the identification of the exit and the successful room evacuation.



It is our aim to explore the interplay between the information overload and coordination costs as mechanisms that can explain the outgoing flux and the evacuation time of pedestrians. Information overload increases with the cluster or group size in which the information is shared and it decreases with the use of heuristics (e.g., identifying a wall that could lead to an exit), while coordination costs increase with the complexity of the system (described by the population size available in the room) and the presence of an obstacle in the room. We investigate the outgoing flux under various population sizes (for each simulated population size, population remains stable during the simulation) in order to capture the interplay between information overload and coordination costs. Moreover, we investigate the evacuation time in order to single out the role of group size (cluster size), interaction with the wall (as decision heuristic) and presence of an obstacle and further clarify the results of the outgoing flux.

\subsection{Structure of the paper} 
\par\noindent
The aim of the paper is stated in Subsection \ref{s:intro}. In Section \ref{s:measure}, we derive formally a balance equation for group--structured mass measures designed to describe the group formation (coagulation and fragmentation) similarly to a "thinking" flow of polymeric chains; see e.g. \cite{Collet} for a nice brief introduction to models like Becker--D\"oring and Smoluchovski for polymeric flows.  Agglomerating and binding monomers leads to our first "group concept". Moving to a totally different level of mathematical description, Section \ref{s:lattice} contains a variation of a simple random walk on a square lattice to describe the interplay between the population size and interpersonal attraction parameters. The stochastic dynamics is biased by the present of a threshold -- our second attempt of a "group concept".  The lattice model is getting to action in Section \ref{s:simulation} and captures the effects of information overload and coordination in large disoriented populations. We conclude the paper with Section \ref{s:conclusions}, where we summarize our main findings concerning our second concept of group of pedestrians.

\section{A measure--theoretical modeling strategy}
\label{s:measure}
\par\noindent
For modeling population dynamics, a natural mathematical tool
is the concept of {\em mass} measure. This measure can flexibly represent either the distribution of individuals in a corridor, their number or their mass localized at a certain location
as a function of possible features. Signed measures form a Banach space. The mass (finite) measures, which are responsible for handling the information pool, are a closed subset in this Banach space. Much of the theory of ordinary differential equations applies directly to dynamics in Banach spaces and therefore to measure--valued differential equations; see \cite{language} for measure--valued equations offering descriptions of the evolution of an unstructured language or our recent work on crowd dynamics \cite{EHM15,EHM16,playing}, where the mass measures evolve in a bounded environment where boundary interactions are allowed. A nice account of measure--theoretical models in the context of crowd dynamics and traffic flows can  be found in \cite{Cristiani}. We need to mention already at this stage that the infinite dimensional geometry is often
counter--intuitive and always requires a  careful mathematical treatment (compare \cite{Bogachev,language}), not only what concerns solvability issues, but also from the perspective of a proper numerical approximation of the evolving measures \cite{Carrillo}. We do not touch here any mathematical concerns around the obtained model. Instead we give a brief derivation of a structured--crowd 
model suitable for dynamics where visibility is obscured. The mathematical analysis needed to ensure the well--posedness of this model will be handled elsewhere. The result of our modeling exercise will turn to be a special case of a 
multi--feature continuity equation;
cf. for instance \cite{MB,BH} or the related setting in \cite{Gyllenberg}.  We intend to provide  here a general modeling framework for building 
``features"--based continuity--like equations for size--ordered interacting populations, which should be seen as an alternative of working with local balance laws using densities. The presentation recalls from \cite{playing}.   In this context, the "location in the corridor" and the "group size" constitute the two "features"; we refer the reader to  \cite{Odo} for a related reading. 
The first feature is a continuous variable, while the second is a
discrete variable. Their evolution is supposed to be continuous in time. In what follows, we  derive a population--balance equation able to describe the evolution of pedestrian groups in
obscured regions, relying on the same principles as in \cite{BH}.

Fix $N\in%
\mathbb{N}
$, let $\Omega\subset%
\mathbb{R}
^{2}$ be the dark corridor (open, bounded with Lipschitz boundary), $S$ - the
observation time interval and $K_{d}:=\left\{  0,1,2,3,..,N\right\}  $ - the
collection of all admissible group sizes. We say that a $Y$ belongs to
$K^{\prime}\subseteq K,$ if it belongs as part of some group with a size \ K.
Furthermore, $\mathfrak{A}_{\Omega}:=\mathfrak{B}^{2}(\Omega)$, $\mathfrak{A}%
_{S}:=\mathfrak{B}^{1}(S)$ are the corresponding Borel $\sigma$--algebras with
the corresponding Lebesgue--Borel measures $\lambda_{x}:=\lambda^{2}$ and
$\lambda_{t}:=\lambda^{1},$ respectively; $\mathfrak{A}_{K_{d}}:=\mathfrak{p}(K_{d})$
is equipped with the counting measure $\lambda_{c}^{\prime}(K):=\left\vert
  K^{\prime}\right\vert $. We call $\lambda_{tx}:=\lambda_{t}\otimes\lambda_{x}$
the \textit{space--time measure} and set $\lambda_{txc}:=\lambda_{t}%
\otimes\lambda_{x}\otimes\lambda_{c_{j}},$ $\mathfrak{A}_{\Omega K_{d}%
}:=\mathfrak{A}_{\Omega}\otimes\mathfrak{A}_{K_{d}},$ $\mathfrak{A}_{S\Omega
  K_{d}}:=\mathfrak{A}_{S}\otimes\mathfrak{A}_{\Omega}\otimes\mathfrak{A}%
_{K_{d}}.$

\subsection{Balance of mass measures}
\par\noindent
Fix $t\in S,$ let $\Omega^{\prime}\in\mathfrak{A}_{\Omega},$ $K^{\prime}%
\in\mathfrak{A}_{K_{d}},$ $S^{\prime}\in\mathfrak{A}_{S}$, we introduce the  target {\em amount} quantity%
\begin{equation}%
  \begin{array}
    [c]{c}%
    \mu_{Y}(t,\Omega^{\prime}\times K^{\prime}):=\text{ number of }Y^{\prime
    }s\text{ present in }\Omega^{\prime}\text{ }\\
    \text{at time }t\text{ and belonging to the group }K^{\prime}\\
  \end{array}
  \label{A10}%
\end{equation}
together with two \textit{inner production} quantities
\begin{equation}%
  \begin{array}
    [c]{c}%
    \mu_{PY\pm}(S^{\prime}\times\Omega^{\prime}\times K^{\prime}):=\text{ number
      of }Y^{\prime}s\text{ which are added }\\
    \text{to (subtracted from) }\Omega^{\prime}\times K^{\prime}\text{ during
    }S^{\prime}\text{ and}\\
    \mu_{PY}=\mu_{PY+}-\mu_{PY-}.
  \end{array}
  \label{A15}%
\end{equation}
Note that these numbers might be non--integers, which is not {\em per se} realistic. Furthermore, group size changes are allowed and we state that a group size 1 exist, which is virtually incorrect. From a sociologic perspective, one is an individual/agent, two is a dyad and from 3 onward we can speak about groups. Consequently, when we describe the change from the size N=2 to N-1=1, the dyad becomes an individual. An increase from 1 to 2 is a transition from individual to dyad in a location, while a change from 2 to 3 is a transition from a dyad to a group. 

Given the nature of the problems we are dealing with, we postulate the following key properties our mass measures are supposed to fulfil:

\begin{description}
\item (P1) \ \ For all $K^{\prime}\in$ $\mathfrak{A}_{K_{d}},$ $\Omega
  ^{\prime}\in$ $\mathfrak{A}_{\Omega}$ $\ $and $t\in S:\mu_{Y}(t,\cdot\times
  K^{\prime})$ and $\mu_{Y}(t,\Omega^{\prime}\times\cdot)$ are measures on their
  respective $\sigma$--algebras $\mathfrak{A}_{\Omega}$ and $\mathfrak{A}_{K_{d}%
  },$ respectively.

\item (P2) $\ \ \mu_{PY\pm}(S^{\prime}\times\Omega^{\prime}\times\cdot),$
  $\mu_{PY\pm}(S^{\prime}\times\cdot\times K^{\prime})$ and $\mu_{PY\pm}%
  (\cdot\times\Omega^{\prime}\times K^{\prime})$ are measures on their
  respective $\sigma$--algebras.
\end{description}

Now, we are in the position to formulate the balance principle:
\begin{equation}%
  \begin{array}
    [c]{c}
    \mu_{Y}(t+h,\Omega^{\prime}\times K^{\prime})-\mu_{Y}(t,\Omega^{\prime}\times
    K^{\prime})=\mu_{PY}(S^{\prime}\times\Omega^{\prime}\times K^{\prime})\\
    \text{for all }t,t+h\in S,\text{ }\Omega^{\prime}\times K^{\prime}%
    \in\mathfrak{A}_{\Omega}\times\mathfrak{A}_{K_{d}},\text{ }S^{\prime
    }:=(t,t+h].\\
  \end{array}
  \label{A20}%
\end{equation}
Addition to $\Omega^{\prime}\times K^{\prime}$, modeled by $\mu_{PY+},$ can
happen by addition \textit{inside} of $\Omega^{\prime}\times K^{\prime}$ as
well as by fluxes \textit{into} $\Omega^{\prime}\times K^{\prime}.$ A similar
remark applies to subtraction and $\mu_{PY-}$. This gives rise to the assumption that
$\mu_{PY+}$ is actually the sum of an interior production part, $\mu_{PY+}^{int},$
and a flux part, $\mu_{PY+}^{flux}$, i.e. an accumulation of interior and boundary productions.  We proceed similarly with $\mu_{PY-}$. Finally, we obtain the representation of the net productions
\[
\mu_{PY}^{int}:=\mu_{PY+}^{int}-\mu_{PY-}^{int}\text{
  and \ }\mu_{PY}^{flux}:=\mu_{PY+}^{flux}-\mu_{PY-}^{flux},
\]%
\begin{equation}
  \ \ \mu_{PY}=\mu_{PY}^{int}+\mu_{PY}^{flux}=\left(  \mu_{PY+}^{int}-\mu
    _{PY-}^{int}\right)  +\left(  \mu_{PY+}^{flux}-\mu_{PY-}^{flux}\right)  .
  \label{A25}%
\end{equation}
We extend $\mu_{Y}(t,\cdot\times\cdot)$ and $\mu_{PY\pm}(\cdot\times
\cdot\times\cdot)$ by the usual procedure to measures $\overline{\mu}_{Y}%
=\mu_{Y}(t,\cdot)$ and $\overline{\mu}_{PY\pm}=\overline{\mu}_{PY\pm}(\cdot)$
on the product algebras $\mathfrak{A}_{\Omega}\otimes\mathfrak{A}_{K_{d}}$ and
$\mathfrak{A}_{S}\otimes\mathfrak{A}_{\Omega}\otimes\mathfrak{A}_{K_{d}},$ respectively.
 Note that the quantities in (P1) and (P2) and the extensions are finite.

The following postulate prevents mass accumulation on sets of measure zero. It
reads as

\begin{description}
\item (P3) \ $\overline{\mu}_{Y}(t,\cdot)\ll\lambda_{xc}$ (absolutely continuous).
\end{description}

Therefore, for all $t\in S$ there exist integrable Radon--Nikodym densities
$u(t,\cdot)=\frac{d\overline{\mu}_{Y}(t,\cdot)}{d\lambda_{xc}},$ i.e.%
\begin{equation}
  \overline{\mu}_{Y}(t,Q^{\prime})=\int_{Q^{\prime}}u(t,(x,i))d\lambda
  _{xc}\text{ \ for all }Q^{\prime}\in\mathfrak{A}_{\Omega K}. \label{A27}%
\end{equation}

The absolute--continuity assumption
\begin{description}
\item (P4) \ $\overline{\mu}_{PY}^{int}\ll\lambda_{txc}$
\end{description}
excludes the presence of $Y^{\prime}s$ on sets of $\lambda_{txc_{1}}$--measure
zero. Furthermore, it assures the existence of the Radon--Nikodym density
\begin{equation}
  f_{PY}^{int}:=\frac{d\overline{\mu}_{PY}}{d\lambda_{txc}}\in L_{loc}%
  ^{1}(S\times\Omega\times K_{d},\mathfrak{A}_{S\Omega K_{d}},\lambda_{txc}).
  \label{A30}%
\end{equation}
To get a reasonable idea for a representation of the flux measure, we
consider the special case $Q^{\prime}=\Omega^{\prime}\times K^{\prime}$ with,
say, $K^{\prime}=\left\{  a,a+1,...,b\right\}  \in\mathfrak{p}(K_{d}).$ The
"surface"
\[
\mathfrak{F}:=\Omega^{\prime}\times\left\{  a\right\}  \cup\Omega^{\prime
}\times\left\{  b\right\}  \cup\partial\Omega^{\prime}\times K^{\prime}%
\]
is the location of any interaction with the outside of $Q^{\prime}$. There are
two locations on $\mathfrak{F}$ to enter or to leave $Q^{\prime}$ from the
outside -- one via $\mathfrak{F}_{1}:=\Omega^{\prime}\times\left\{  a\right\}
\cup\Omega^{\prime}\times\left\{  b\right\}  ,$ the other one through
$\mathfrak{F}_{2}:=\partial\Omega^{\prime}\times K^{\prime}$.


The unit--outward normal field $\mathbf{n}=\mathbf{n}(x,\kappa)$ on
$\mathfrak{F}$ can be split into two orthogonal components, $\mathbf{n}%
=\mathbf{n}_{x}+\mathbf{n}_{\kappa},$ $\mathbf{n}_{x}=(n_{x},0),$
$\mathbf{n}_{\kappa}=(0,n_{\kappa}),$ respectively. It is $n_{\kappa}(x,a)=-1$,
$n_{\kappa}(x,b)=+1$ and $n_{x}=n_{x}(x,\kappa)$ is the a.e. existing outward
normal on $\partial\Omega.$ 
Borrowing from the theory of Cauchy interactions, cf. \cite{Schuricht}, e.g., we postulate the following:

\begin{description}
\item (P5) Assume, for all $t\in S$, the existence of two {\em flux vector fields}
  \begin{eqnarray}
    j_{x}(t,\cdot)&:&\Omega\times K_{d}\rightarrow
    \mathbb{R}^2\nonumber\\
    j_{\kappa}&:&\Omega\times K_{d}\rightarrow \mathbb{R}\nonumber
  \end{eqnarray}
  \end{description}
  inter-linked via 
  \[
  \overline{\mu}_{PY}^{flux}:=\overline{\mu}_{PYx}^{flux}+\overline{\mu
  }_{PY\kappa}^{flux}\text{,}%
  \]
  where
  \begin{equation}%
    \begin{array}
      [c]{rl}%
      \text{ }\\
      \overline{\mu}_{PYx}^{flux}(S^{\prime}\times\Omega^{\prime}\times K^{\prime
      })&:=\int_{S^{\prime}}\int_{\mathfrak{F}_{_{2}}}-j_{x}(\tau,x,i)\cdot
      n_{x}(x,i)d\sigma_{x}d\lambda_{c}d\tau\text{ \ }\\
      &\text{and}\\
      \overline{\mu}_{PY\kappa}^{flux}(S^{\prime}\times\Omega^{\prime}\times
      K^{\prime})
      &:=\int_{S^{\prime}}\int_{\Omega^{\prime}}-j_{\kappa}(\tau,x,b)n_{\kappa
      }(x,b)\\&-j_{\kappa}(\tau,x,a)n_{\kappa}(x,a)dxd\tau.\\
    \end{array}
    \label{A35}%
  \end{equation}
In (\ref{A35}),  $\sigma_{x}$ - is the $1D$--curve length measure. The term $\overline{\mu
}_{PYx}^{flux}(S^{\prime}\times\Omega^{\prime}\times K^{\prime})$ calculates
the net gain/loss of the $Y^{\prime}s$ in $\Omega^{\prime}$ belonging to one
of the size groups from $K^{\prime}$ due to physical motion from/to the
outside of $\Omega^{\prime}$ into/out of $\Omega^{\prime}$. Note that the quantity $\overline{\mu}_{PY\kappa}^{flux}(S^{\prime}\times\Omega^{\prime}\times\left\{  i\right\}
)$ calculates the net gain/loss of the $Y^{\prime}s$ in $\Omega^{\prime}$
belonging to the size group labelled by $i$. 
There is no interaction with groups of
size $\kappa>N$ or $\kappa<0$ since these  sizes are considered here as unadmissible. In this case, we require%
\begin{equation}
  j_{\kappa}(t,x,0)=j_{\kappa}(t,x,N)=0\text{ \ for all }t\in S,\text{ }%
  x\in\Omega.\label{A37}%
\end{equation}
Introducing the \textit{discrete partial derivative} by
\[
\text{ }\partial_{i}^{d}j_{\kappa}(t,x,i):=j_{\kappa}(t,x,i+1)-j_{\kappa
}(t,x,i),\text{ }i\in K
\]
and assuming $u,$ $f_{PY}^{int},$ $\operatorname{div}_{x}j_{x}$ and
$\partial_{\kappa}^{d}j_{\kappa}$ to be sufficiently regular, we obtain%

\begin{align*}
  \overline{\mu}_{PY}^{flux}(S^{\prime}\times Q^{\prime}) &  =\int_{S^{\prime}%
  }\int_{Q^{\prime}}-\operatorname{div}_{x}(j_{x}(\tau,x,i)d\lambda_{c}dxd\tau\\
  &  +\int_{S^{\prime}}\int_{Q^{\prime}}-\partial_{i}^{d}j_{\kappa}(t,x,i)d\tau
  dxd\lambda_{c}.
\end{align*}
Combining (\ref{A20}) -- (\ref{A35}), Fubini's theorem, and division by $h>0$, imply%
\[%
\begin{array}
  [c]{c}%
  \int_{Q^{\prime}}\frac{1}{h}\left(  u(t+h,x,i)-u(t,x,i)\right)  dxd\kappa\\
  =\int_{Q^{\prime}}\frac{1}{h}\int_{t}^{t+h}\left(  f_{PY}^{int}(\tau
    ,x,i)-\left(  \operatorname{div}_{x}j_{x}(\tau,x,i)+\partial_{i}^{d}j_{\kappa
      }(t,x,i)\right)  \right)  d\tau dxd\lambda_{c}.\\
\end{array}
\]
Under appropriate smoothness conditions on $u,$ $f_{PY}^{int},$ $j_{x}$ and
$j_{\kappa\text{ \ }}$we obtain in the limit $h\rightarrow0$ (the classical
continuity equation with a slightly different interpretation of the entries)%

\begin{equation}
  \frac{\partial u}{\partial t}(t,x,i)+\left(  \operatorname{div}_{x}%
    j_{x}+\partial_{i}^{d}j_{\kappa}(t,x,i)\right)  =f_{PY}^{int}(t,x,i).
  \label{A51}%
\end{equation}
The existence of the underlying measures is a matter that will be investigated at a later stage, in a different context.
 
As in \cite{playing}, we can point out a direct connection with a suitable Becker--D\"oring--like  model for colloidal interactions (cf. e.g. \cite{Collet} and references cited therein), which mimics well the dynamics of groups of individuals in dark. 
To do so, one has to specify the flux vectors
$j_{x}$ and $j_{\kappa}$ as well as the production term $f_{PY}^{int}.$ 

For $i=1,...,N$, we set  $$J_{x}(t,x,i)=-D_{i}(u_{i}(t,x)) {\rm grad}_{x}(u_{i}(t,x)),$$
$$f_{PY}^{int}(t,x,i)=
\left\{ \begin{array}{lr}
    \sum\nolimits_{i=1}^{N}
    \alpha_{i}u_{i}-\sum\nolimits_{i=1}^{N}
    +\beta_{i}u_{i}u_{1} & \mbox{ if }   i=1,\\
    \beta_{i-1}u_{i-1}u_{1}-\beta_{i}u_{i}u_{1} & \mbox{ if }  i\in\{2,...,N-1\},\\
    \beta_{N}u_{N-1}u_{1} & \mbox{ if }  i=N,\\
  \end{array}
\right.
.$$
The discrete
derivative $j_{\kappa}(t,x,i)$  corresponds to
\begin{eqnarray}
  j_{\kappa
  }(t,x,i)& =& -\alpha_{i}u_{i}(t,x), \ i=1,2,...,N-1,\nonumber\\
  j_{\kappa}(t,x,0)& =& j_{\kappa}(t,x,N)=0.\nonumber
\end{eqnarray}

Specifying $j_{x}(t,x,i)$ as some sort of a degenerating diffusion flux in the manner above
means: Individual groups of size $i$ recognize whether a group of the same
size is in their immediate neighborhood and they tend to avoid moving into
the direction of such groups. Employing a Fickian law seems to be the simplest
way to model this.   The degeneration in the flux appears especially if the dynamics is thought to take place in domains with obstacles, where the percolation of the "thinking" fluid locally clogs when obstacles are too close from each other.  

The production term $f_{PY}^{int}$ models merging interactions between
groups of size $i\in K$ and "groups" of size $i=1$: If an individual meets a group of size $i<N$, then it might happen, that this
single merges with the group. This turns the group into a group of size $i+1$
and leads to a "gain" for groups of size $i+1$ (modeled by $+\beta_{i}%
u_{i}u_{1}$) and a loss for groups of size $i$ (modeled by $-\beta_{i}%
u_{i}u_{1}$). In any such joining situation the group with $i=1$ looses members
(modeled by $-%
{\textstyle\sum\nolimits_{i=2}^{N}}
\beta_{i}u_{i}u_{1}$). Note, that this model allows only for direct
interaction between groups of size $i$ with groups of size $1$. Using a Smoluchowski dynamics instead would relax this constraint \cite{Collet}.  The $\alpha
$--terms model the group fragmentation effect: It might happen, that an individual
leaves a group of size $i\geq2$. This leads to a loss for the groups of size
$i$ (modeled by $-\alpha_{i}u_{i}$) as well as to a gain for the groups of size $i-1$ and
a gain for the groups with $i=1$ (modeled by $%
{\textstyle\sum\nolimits_{i=2}^{N}}
\alpha_{i}u_{i}$). In simple crowd simulations \cite{playing}, the coefficients $\alpha_{i},$ $\beta_{i}\geq0$ and $D_{i}>0$ can be assumed to be constant.
It is interesting to note that the fragmentation terms express a flux rather
than a volume source or sink. In the same way as aging can be seen as a flux
("people change their age group by aging with (speed 1)" ) $Y$'s change their
size group by "degradation" of their group. Nevertheless, for a fixed size choice $i$,  the
expressions $\alpha_{i}u_{i}$ and $\alpha_{i-1}u_{i-1}$ still remain "volume
sources" and "volume sinks", respectively. 

The mathematical analysis of this model can be approached for instance by adapting the techniques from \cite{DaPrato} to our setting combined with arguments specific to handling porous media--like equations.   
However, we do not expand on this matter here and prefer instead to propose a lattice model not only to describe at the microscopic level the motion of the crowd in dark, but also to quantify the information overload  and coordination in crowded populations. 

\vfill\eject
\section{A lattice--based modelling strategy}
\label{s:lattice}

\subsection{The lattice model}
\label{s:l-model}
\par\noindent
Our model was introduced in \cite{cm2012,cm2013} by adapting 
ideas introduced in \cite{abc2011} to study the ionic currents 
in cell membranes and further developed in \cite{abc2014}. For an introduction to stochastic processes, we refer the reader e.g. to \cite{Pavliotis}.

To describe our model, we start off with the construction of the lattice.  
Let $e_1=(1,0)$ and $e_2=(0,1)$ denote the coordinate vectors in 
$\mathbb{R}^2$.
Let $\Lambda\subset\mathbb{Z}^2$ be a finite square with 
odd side length $L$. We refer to this as  the {\em corridor}. 
Each element $x$ of $\Lambda$ will be called a {\em cell} or {\em site}.
The external boundary of the corridor is made of four segments 
made of $L$ cells each; 
the point at the center of one of these four sides is called {\em exit}.

Let $N$ be  positive integer denoting the (total) 
{\em number of individuals} inside 
the corridor $\Lambda$. 
We consider the state space $X=\{0,\dots,N\}^\Lambda$.  
For any state $n\in X$, we denote by 
$n(x)$ be the {\em number of 
individuals} at cell $x$.  

We define a Markov chain $n_t$ on the finite state space 
$X$ with discrete time $t=0,1,\dots$.
The parameters of the process are the integer 
$T\ge0$ called {\em threshold}, $W\geq 0$ called \emph{wall stickiness},
and the real number $R\in[0,1]$ called the {\em rest parameter}.
We finally define the function $S:\mathbb{N}\to\mathbb{N}$ 
such that 
\begin{displaymath}
S(k)= 1  \textrm{ if } k>T
\;\;\textrm{ and }\;\;\;
S(k)=k+1  \textrm{ if } k\le T
\end{displaymath}
for any $k\in\mathbb{N}$. Note that for $k=0$ we have $S(0)=1$.

At each time $t$, the $N$ individuals move simultaneously within the corridor 
according to the following rule:
for any cell $x$ situated in the interior of the corridor $\Lambda$, and all 
$y$ nearest neighbor of $x$,
with  $n\in X$, we define the weights
$w(x,x)=RS(n(x))$
and
$w(x,y)=S(n(y))$.
Also, we obtain the associated probabilities 
$p(x,x)$ and  $p(x,y)$
by dividing the weight by the normalization 
\begin{displaymath}
w(x,x)+
\sum_{i=1}^2 w(x,x+e_i)+
\sum_{i=1}^2 w(x,x-e_i).
\end{displaymath}
Let now $x$ be in one of the four corners of the corridor $\Lambda$,  
and take
$y$ as one of the two nearest neighbors of $x$ inside $\Lambda$. 
For $n\in X$, we define the weights
$w(x,x)=RS(n(x))+2 W$
and
$w(x,y)=S(n(y))+W$
and the associated probabilities $p(x,x)$ and $p(x,y)$ obtained 
by dividing the weight by the suitable\footnote{To get a probability, we divide the weights
by  the sum of the involved weights: It is about 3 terms if the 
site lies in a corner of the corridor, 4 terms if it is close to a border, and then 
5 terms for a pedestrian located in the core. } normalization.
In this case we are introducing the additional term $W$ that is mimicking the stickiness of the wall. 
That  is this parameter takes into account the possibility that people prefer to move nearby the wall in condition of lack of visibility. 

It is worth stressing, here, 
that $T$ is not a threshold in $n(x)$,
the number of individuals per cell, but 
a threshold in the probability that such a cell is likely 
to be occupied or not.  
 
For $x\in \Lambda$ neighboring the boundary (but neither in the corners, nor  
neighboring the exit), 
$y$ one of the two nearest neighbor of $x$ inside $\Lambda$ and 
neighboring the boundary,
$z$ the nearest neighbor of $x$ in the interior of $\Lambda$,
and $n\in X$, we define the weights
$w(x,x)=RS(n(x))+W$, 
$w(x,y)=S(n(y))+W$, 
and
$w(x,z)=S(n(z))$.
The associated probabilities $p(x,x)$, $p(x,y)$, and $p(x,z)$ are obtained 
by dividing the weight by the suitable normalization. 

Finally, we define the weights if $x$ is the site in 
$\Lambda$ neighboring the exit.
We propose two different choices mimicking two different situations.
If the exit is clearly identifiable, particles in the neighboring 
cell exit with probability $1$. This exit modelling 
will be called \emph{sure exit} case.
If the exit is not clearly identifiable, we altough assume 
it is the most likely site to jump to and we treat it 
as if it were occupied by the threshold number of pedestrians.
More precisely, 
in this case, called the \emph{threshold exit case}, 
if  
$y$ is one of the two nearest neighbor of $x$ inside $\Lambda$ and 
neighboring the boundary and 
$z$ is the nearest neighbor of $x$ in the interior of $\Lambda$,
we define the weights
$w(x,x)=RS(n(x))$,
$w(x,y)=S(n(y))+W$,
$w(x,z)=S(n(z))$,
and 
$w(x,\textrm{exit})=T+1$,
for the configuration $n\in X$. 
The associated probabilities $p(x,x)$, $p(x,y)$, $p(x,z)$,
and $p(x,\textrm{exit})$ are obtained 
by dividing the weight by the suitable normalization. 

At time zero, the individuals are randomly located inside the corridor.
The dynamics is then defined as follows:
at each time $t$,  the position of all the individuals 
on each cell is updated according to the probabilities defined 
above. If one of the individuals jumps on the exit 
cell a new individual is put on the cell of $\Lambda$ 
neighboring $\partial\Lambda$ 
and opposite to the exit site. 

\subsection{Social groups in search for information}
\par\noindent
As we argued before, information sharing happens in smaller social groups. In our lattice model, group size is modeled as an attraction parameter that stimulates agents to move in particular locations of the lattice in which a predefined number of agents is already present. We will define the thresholds for this attraction parameters using existing research on human groups formation. In an overview of how social cognition emerges in groups \cite{caporael}
it is argued that the dyad (a group of two) is the smallest social entity in which microcoordination processes take place. In a similar vein, 
in \cite{moreland} it is argued that dyads are specific forms of social aggregates in which coordination costs are minimal as these aggregates consist of just a single relationship, are simpler than larger groups, in other words, dyads are social aggregates in which information pool and coordination costs are low. Following the categorization of social aggregates based on size, 
in \cite{caporael} the author 
further distinguishes between work/family groups, demes and macrodemes. 
The small groups (of size 5) that in evolutionary terms resemble small families, are social aggregates in which division of labor can be observed and coordination constraints increase as they are formed of several dyadic relations \cite{moreland}. 
The so--called 
demes (larger groups of size 30 or bands) are the first organizational forms in which economic activities and transactions are undertaken and therefore involve higher coordination. Finally, the macrodemes (larger communities of size 300) are social aggregates that can deal with highly complex environments (including survival under limited resource availability) \cite{caporael}. In other words, the macrodemes are the social aggregates in which both potential information pool and coordination costs are high. 
Although, as argued in \cite{caporael}
the group sizes used to distinguish between these social aggregates are not exact estimates, we decided to follow this categorization in defining the small group in which potential information sharing takes place.
 In the simulation, these various small group sizes described above are represented by the 
number of agents per site that maximizes the probability of an agent to jump on that site.  For this reason, we will use the parameter $T=0, 2, 5, 30,$ and $300$.

We further on consider the interaction with the walls of the room as a decision heuristic used to oversimplify the complexity of the available information. Decision--makers often use heuristics to speed--up their decision processes and as such they select a small amount of information they process. In particular, the wall could signal the likely location of an exit, as an exit is always tied to a wall. We therefore introduce this heuristic as a way of mitigating the information overload, especially in large populations. Finally, we model the effect an obstacle located either at the center of the room, or close to the exit has on the outgoing flux and evacuation time. An obstacle increases the coordination costs as it may interfere with the cluster formation and the effective movement towards the exit.

\section{The lattice model in action -- simulation of social groups}
\label{s:simulation}
\par\noindent
We explore numerically the effect of groups sizes (threshold) on the outgoing flux, on the interaction with the walls and eventual fixed obstacles, as well as on the evacuation time.  

\subsection{The outgoing flux}
\label{s:l-res}
\par\noindent
In this section,  
the main quantity of interest is the outgoing flux
\cite{cm2012,cm2013,abc2011,abc2014}, i.e., the average
number of individuals exiting the corridor per unit of time. 
Recall that the unit of time is the time in which 
all the individuals are moved. 
In our simulations, we consider as corridor $\Lambda$ a 
square lattice with side $L=101$ and $W=0$  (if not mentioned otherwise). 

\begin{figure}[h]
\begin{picture}(400,140)(0,0)
\centering
\put(-6,0){
  \includegraphics[width=0.62\textwidth]{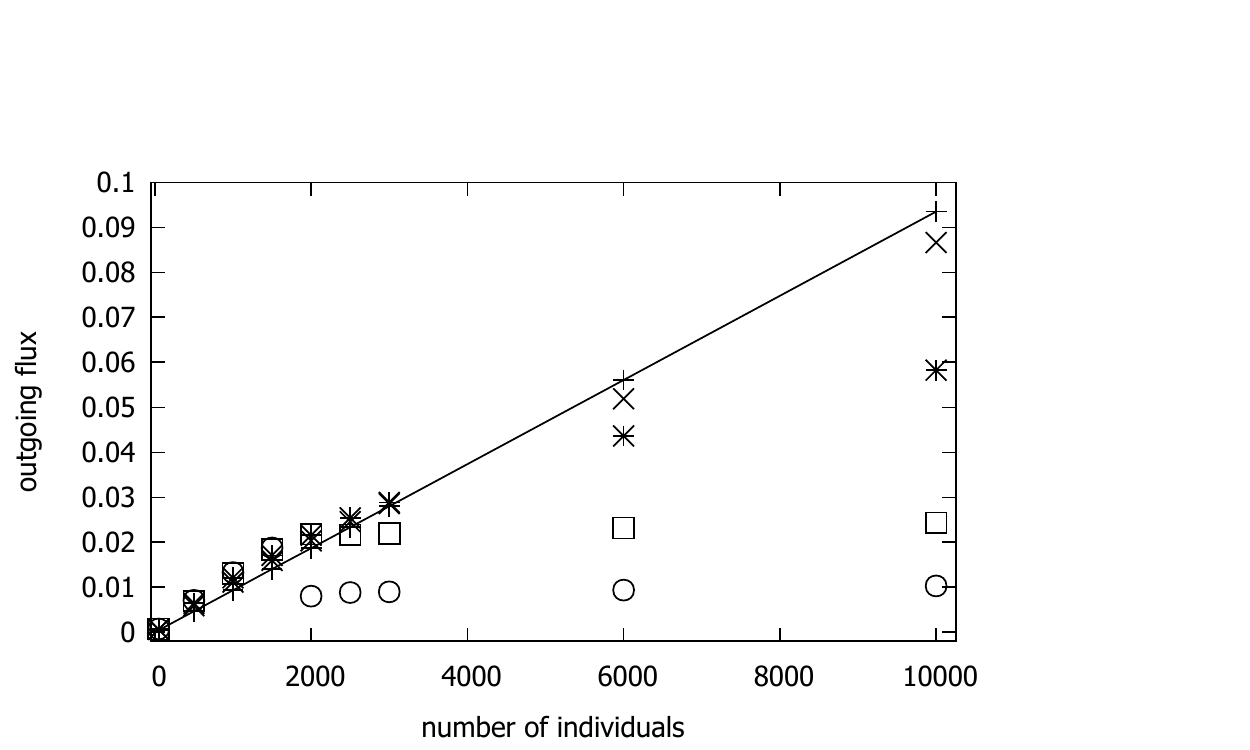}
}
\put(230,0){
  \includegraphics[width=.62\textwidth]{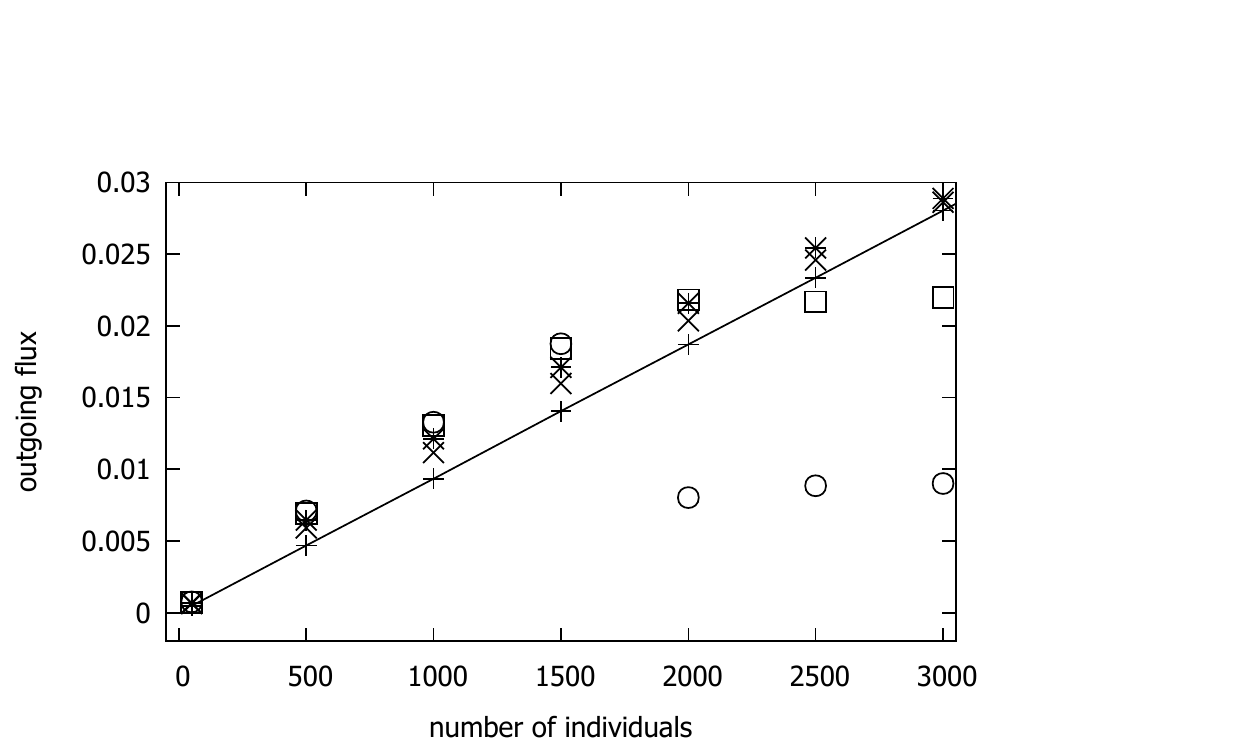}
}
\end{picture}
\caption{{\bf Outgoing flux for $R=0$.}
On the left: outgoing flux for $R=0$ and {\em threshold exit rule}.
On the right: zoom for the outgoing flux for $R=0$ case and $N\leq 3000$.
The symbols
$+$, 
$\times$, 
$*$,
the square and the circle
refer
respectively to the cases $T=0,2,5,30,300$.
}
\label{fig1n}
\end{figure}

\begin{figure}[h]
\begin{picture}(400,140)(0,0)
\centering
\put(-6,0){
  \includegraphics[width=0.62\textwidth]{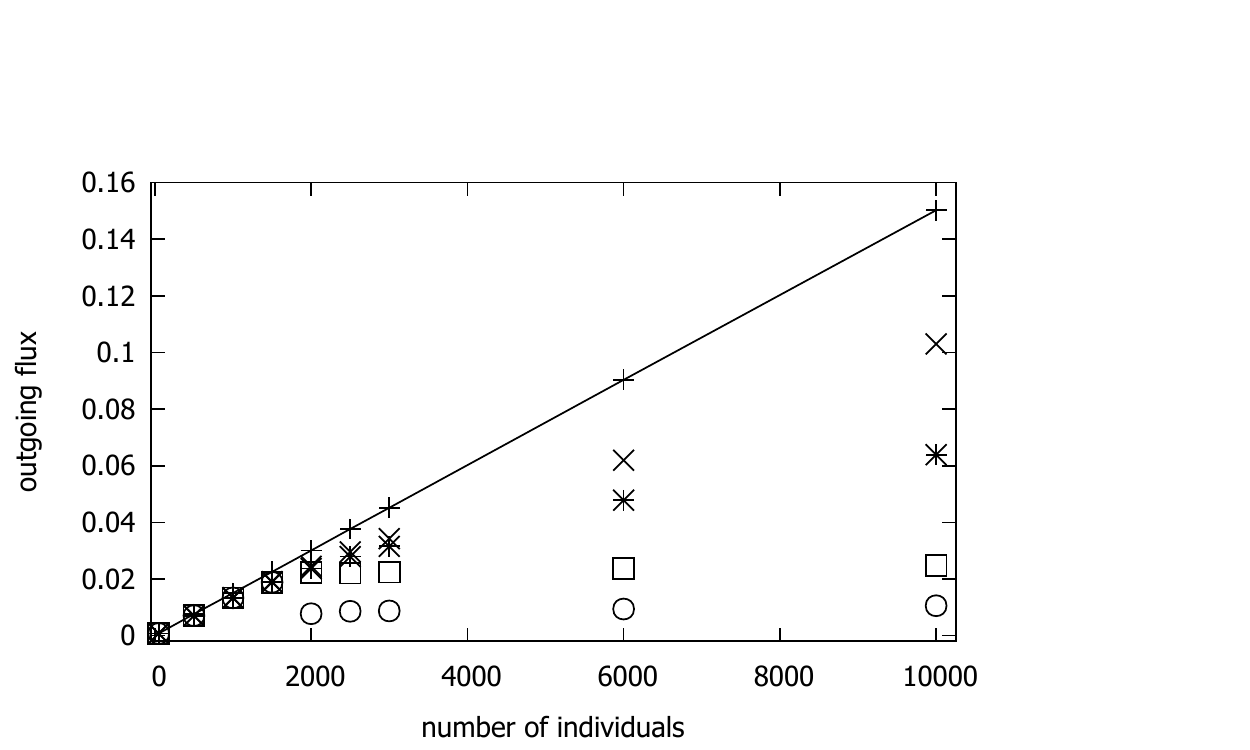}
}
\put(230,0){
  \includegraphics[width=.62\textwidth]{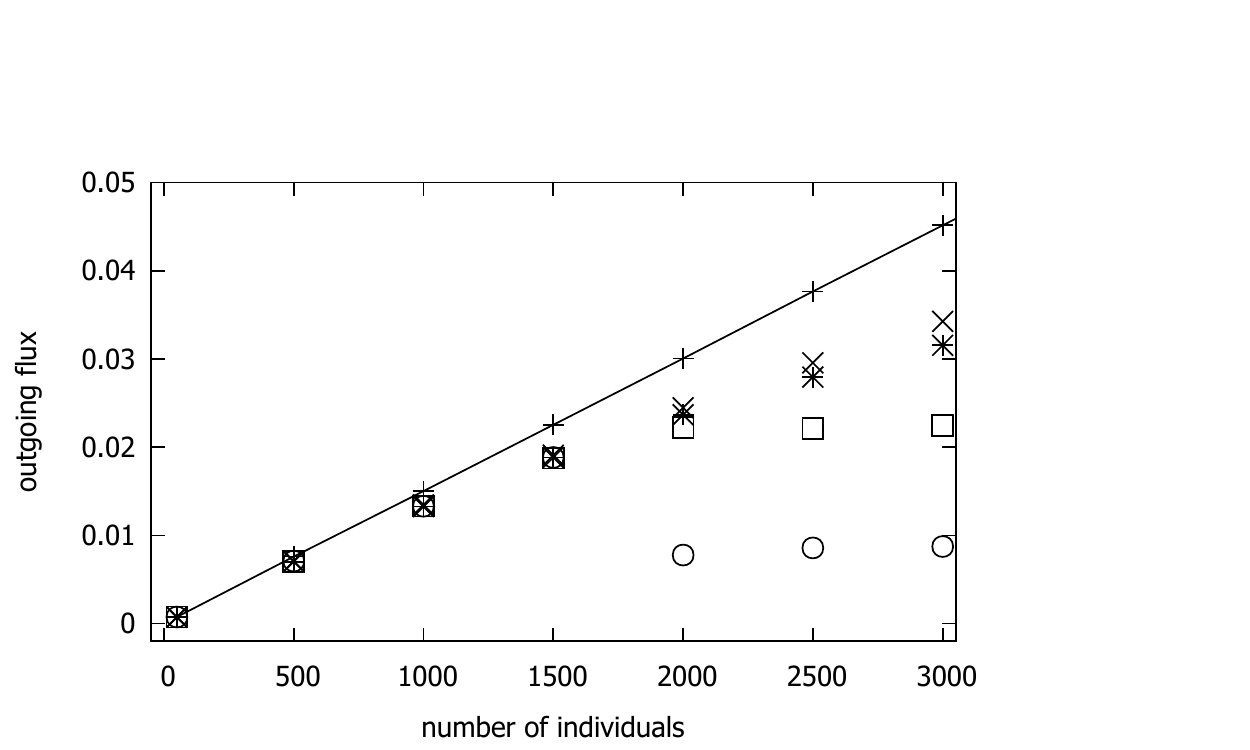}
}
\end{picture}
\caption{{\bf Outgoing flux for $R=0$.}
On the left: outgoing flux for $R=0$ and sure exit rule.
On the right: zoom for the outgoing flux for $R=0$ case and $N\leq 3000$.
The symbols
$+$, 
$\times$, 
$*$,
the square and the circle
refer
respectively to the cases $T=0,2,5,30,300$.
}
\label{fig2n}
\end{figure}

\begin{figure}[h]
\begin{picture}(400,140)(0,0)
\centering
\put(-6,0){
  \includegraphics[width=0.62\textwidth]{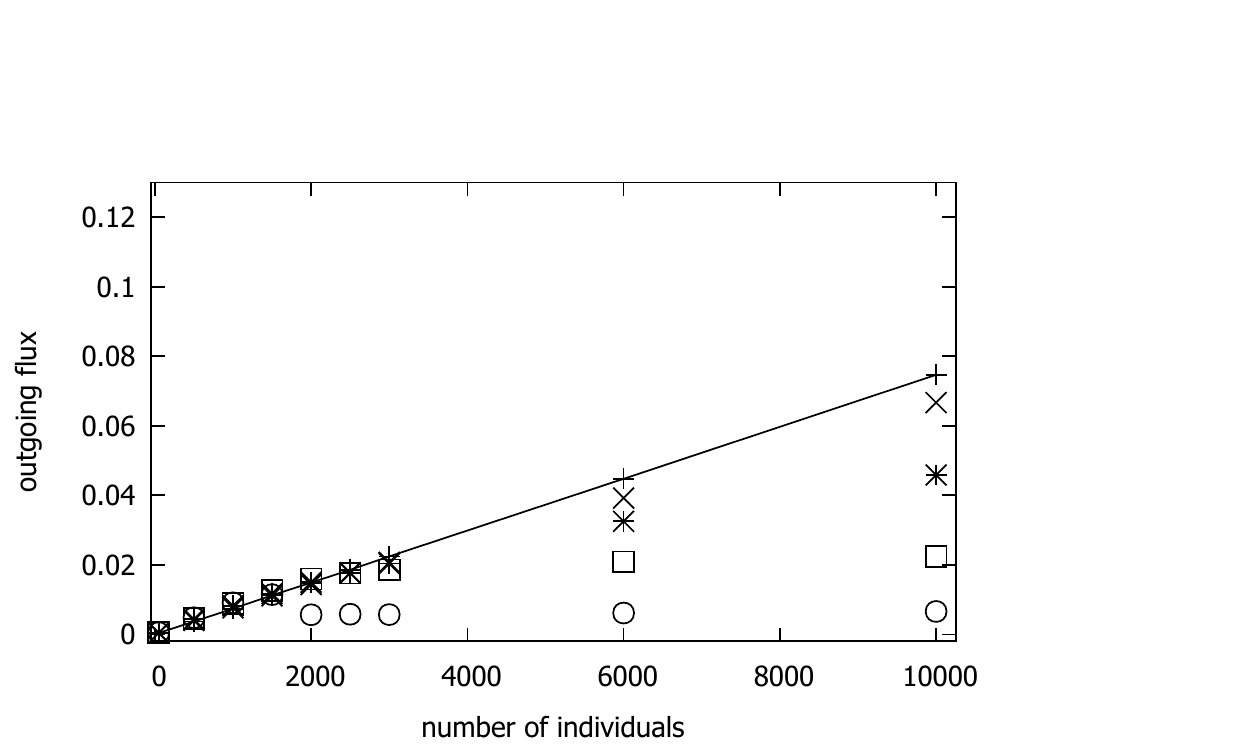}
}
\put(230,0){
  \includegraphics[width=.62\textwidth]{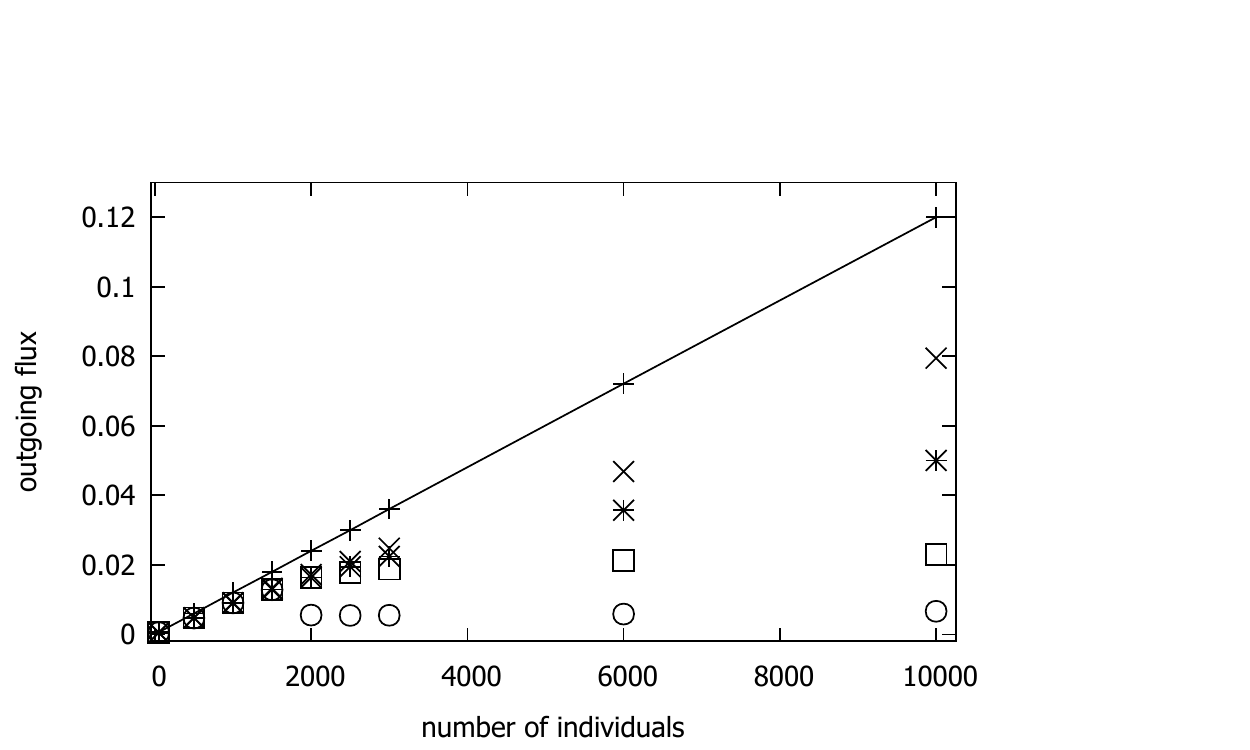}
}
\end{picture}
\caption{{\bf Outgoing flux for $R=1$.}
On the left the plot of the threshold exit case while on the right the plot of the sure exit case.
The symbols
$+$, 
$\times$, 
$*$,
the square and the circle
refer
respectively to the cases $T=0,2,5,30,300$.
}
\label{fig3n}
\end{figure}

Fig.~\ref{fig1n} and Fig. \ref{fig2n} show the outgoing flux plotted as
a function of the number of pedestrians $N$ in the case in
which $R=0$ for the two different exit rules considered, while in  
Fig.~\ref{fig3n} the same for
$R=1$ is plotted.
In Fig.~\ref{fig1n} and Fig. \ref{fig2n}, on the right side,     
it is represented the zoom of the picture on the left side of
Fig.~\ref{fig1n} and Fig. \ref{fig2n}, respectively,
for small population size  (below $3000$).

For a large number of pedestrians the cooperation effects incorporated by the use of the threshold $T$ are not as efficient as a Brownian (egoistic) walk.    
In the case of the threshold exit rule, 
for  a small number of particles (say, below $N=3000$), 
we observe that  the presence of the threshold in our lattice 
dynamics accelerates the particles motion leading to fluxes
higher than the ones obtained for the Brownian motion.
Furthermore, switching from the rest probability $R=1$ to the rest
probability  $R=0$ enhances the threshold influence on the evacuation flux.
This effect is new and non--intuitive. 

\begin{figure}[!ht]
\begin{picture}(300,200)(0,0)
\centering
\put(40,-8){
  \includegraphics[width=.90\textwidth]{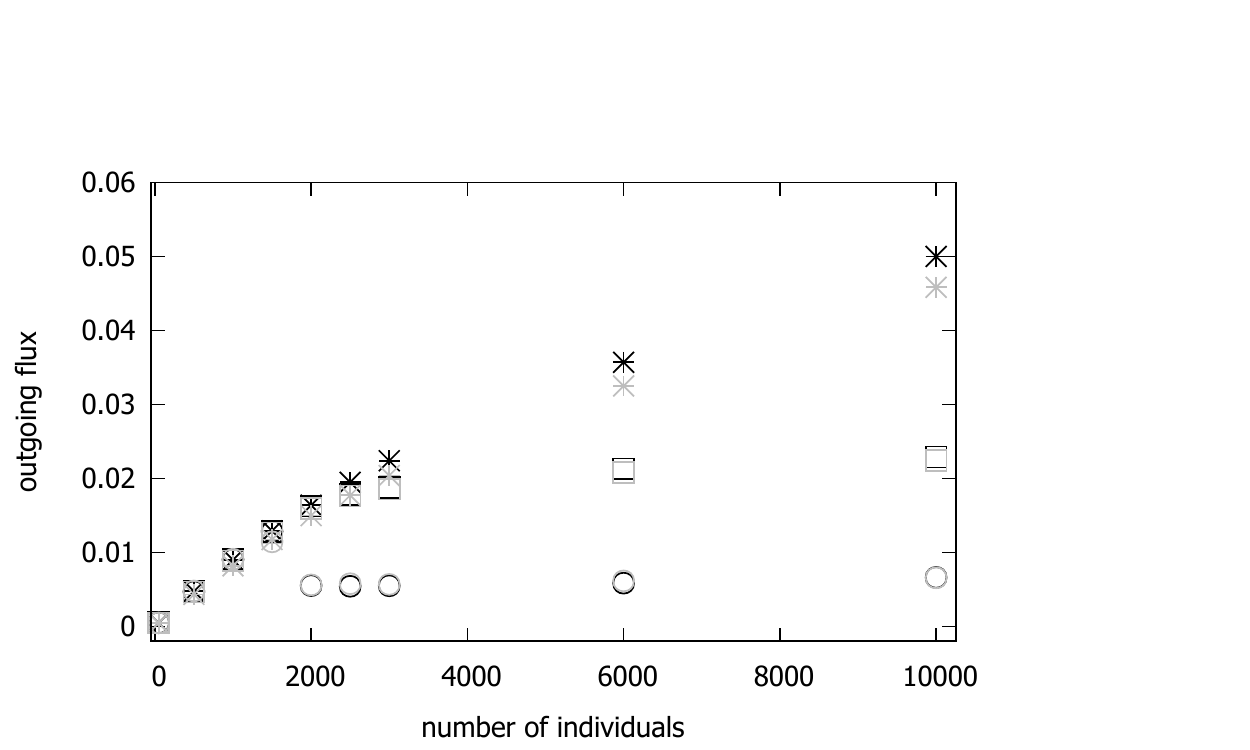}
}
\end{picture}
\caption{{\bf Averaged outgoing flux vs. number of pedestrians.}
 We show a comparison of the simulation results for $T=5$ ($ \ast$), $30$ (squares), $300$ (circles), in the case of \emph{sure exit} (in black), and in the case of \emph{threshold exit} (in gray) for $R=1$. Note that gray and black circles and squares are overlapping.
}\label{fig4n}
\end{figure}

As one expects, the choice of the \emph{sure exit} favors the outgoing  flow. Indeed, in this case the individuals in front of the exit go always out of $\Lambda$ at the first next update of their position, while in the other case they can remain in $\Lambda$ with positive probability.
This is neatly proven by the data for the case of $T=0$, $2$,
but it is not so evident for larger thresholds. 
In Fig.~\ref{fig4n} data relating to the two exit rules are compared 
for $T=5,30,300$.
Black and grey symbols are closer and closer when $T$ is increased, 
this means that the difference between the two cases vanishes 
when $T$ grows.

This effect can be explained as the consequence of our choice for the 
probability to exit for a particle facing the exit in the 
\emph{threshold exit} case. Indeed, for small values of $T$, such as 
$T=0,2$, 
the weight $T+1$ we assign to the exit site is comparable to the weight 
associated with the other 
neighboring sites, so that the particles often remain in the 
system instead of jumping to the exit. 
This is not the case for larger values of $T$, since the 
weight associated with the exit site is likely to be significantly 
larger than the one assigned to the other neighboring sites. 
Hence, in this case, the system behaves similarly to the 
sure exit case.

This remark explains the effect observed in Fig.~\ref{fig1n} and 
stressed in the zoom.
 For small values of $N$, the average outgoing exit flux has an increment in the case of large threshold with respect to the independent random walk case ($T=0$), if we consider the threshold exit rule. 
This is due to the fact that in a regime of small number of individuals, it is really unlikely to have many individuals on the nearest neighbors of the sites closest to the exit, so the weight of the exit site is clearly much larger
  than the other nearest neighbors. 
  Thus, the probability to jump out is significantly bigger for large $T$ than in the case of small values of the threshold.

This phenomenon is observed only in the case of small density of individuals, 
and it is obviously not present for the sure exit rule. 
If the particle density is increased, then the interaction among individuals 
plays a much more important role in the dynamics, influencing 
crucially  the behavior of the overall flux 
and the overshoot finally disappears.

\begin{figure}[h]
\begin{center}
\includegraphics[width=8cm,height=5cm]{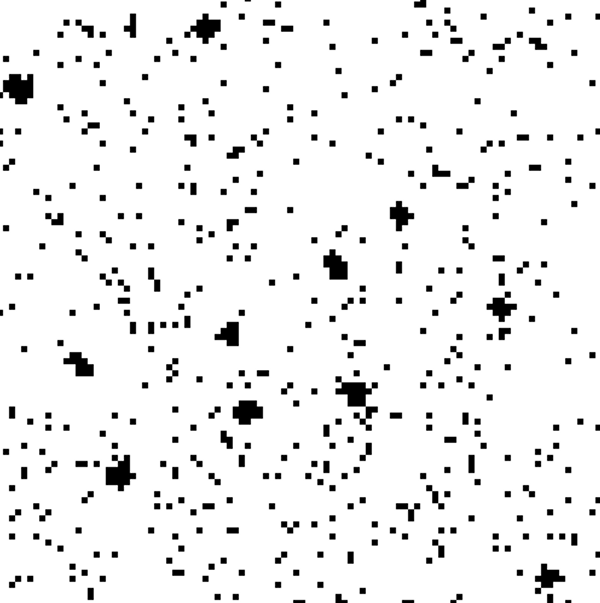}
\end{center}
\caption{{\bf Typical configuration.}
Typical configuration
of the system at large time in the case 
$R=1$, $T=300$, $L=101$, and $N=10000$. 
White and black points denote, respectively, 
empty and occupied sites of the lattice. 
}
\label{fig5n}
\end{figure}

The simulation results show some very interesting 
patterns for the interplay between potential 
information pool and coordination costs on the 
rate of exit per unit of time. 
As the system complexity (illustrated by the population size) 
increases from low to average it seems that the higher potential 
information pool associated with larger subgroups pays off and 
the rate of exit increases above the base rate ($T=0$). 
However, as the complexity of the system further increases 
(above population size $2000$) 
the coordination costs (generated by the fact that very large clusters 
are actually formed) outweigh the benefits of higher 
potential information pool. The rate of exit levels off for 
$T=300$ (at around population size $2000$), 
for $T=30$ at around $2500$, for $T=5$ at around $6000$, 
while for $T=2$ the exit rate is only slightly lower than the 
base rate at the population size $10000$. 

A plausible interpretation of these results is that as environmental complexity increases from average to high, the coordination costs become the most important factor that drives the exit rate per unit of time.
For simpler systems, a high number of cognizers organized in groups seem to be essential in getting as many people out of the room as possible, while for highly complex systems, dyads seem to suffice. 

In other words, in systems with a substantial number of agents 
(higher than $2000$), for $T=300$, 
very large groups can easily form and 
as they are built, coordination costs block effective 
information sharing in these groups and the large clusters 
become stuck in a lengthy decision process. 
Large groups ($300$ members) once formed, start information 
sharing and try to reach a decision on actions to be taken. 
However, integrating the vast amount of information brought 
in by the large number of agents takes a substantial 
amount on time and effort. 

Next to the coordination constraints, members in large groups tend to exert less effort towards the collective goal, 
therefore social loafing increases with groups size 
\cite{ingham}. It is therefore likely that agents will spend a 
substantial amount of ``effortless'' time in such large clusters that 
will ultimately prevent them from exiting as indicated in 
Fig.~\ref{fig5n}
that depicts the cluster formation behavior for $N=10000$ and $T=300$.        

\subsection{The interaction with the wall}
\label{s:l-wal}
\par\noindent
We study the dependence of the dynamics on the possible choice of the parameter $W$, which we refer to as {\em wall stickiness}. 
We introduce this parameter to take into account the possibility that the pedestrian may have the tendency to move close to the wall as main personal reaction to the lack of visibility and a heuristic for finding an exit.
 
We expect that the choice of a positive value for the parameter $W$ can improve the outward flux. This intuition follows from the fact that the number of sites neighboring the boundary are of order of $\mathcal{O}(L)$, 
 while the total number of sites is of order of $\mathcal{O}(L^2)$.
So the choice of the parameter $W>0$ (illustrative of a decision heuristic "A wall indicates the location of an exit") helps reducing the random walk on a number of sites of lower order and the individuals will find the exit earlier.

On the other hand, we expect that the agents localization close to the wall may favor the formation of clusters of particles as all agents may use the same heuristic. In our view, the increased coordination costs associated with this agglomerations close to the walls could be a mechanism leading to the slowing down of the evacuation.

\begin{figure}[!ht]
\begin{picture}(400,140)(0,0)
\centering
\put(-6,0){
  \includegraphics[width=0.62\textwidth]{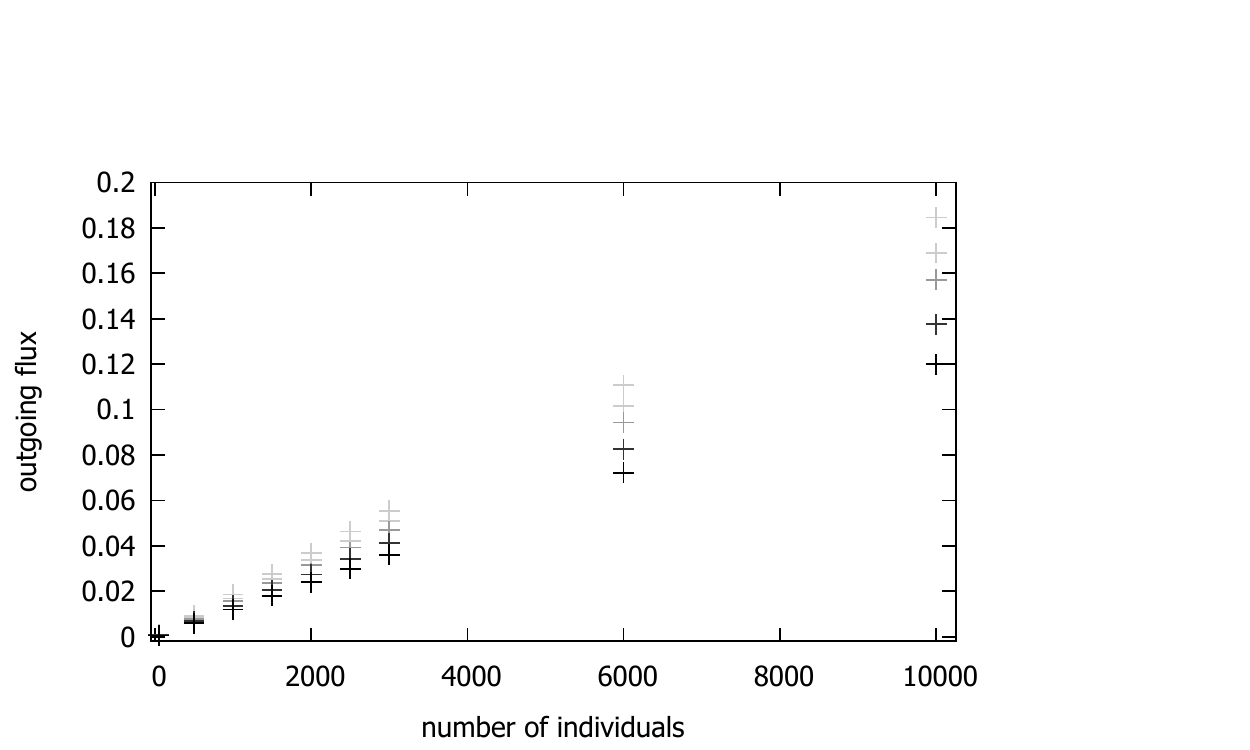}
}
\put(230,0){
  \includegraphics[width=.62\textwidth]{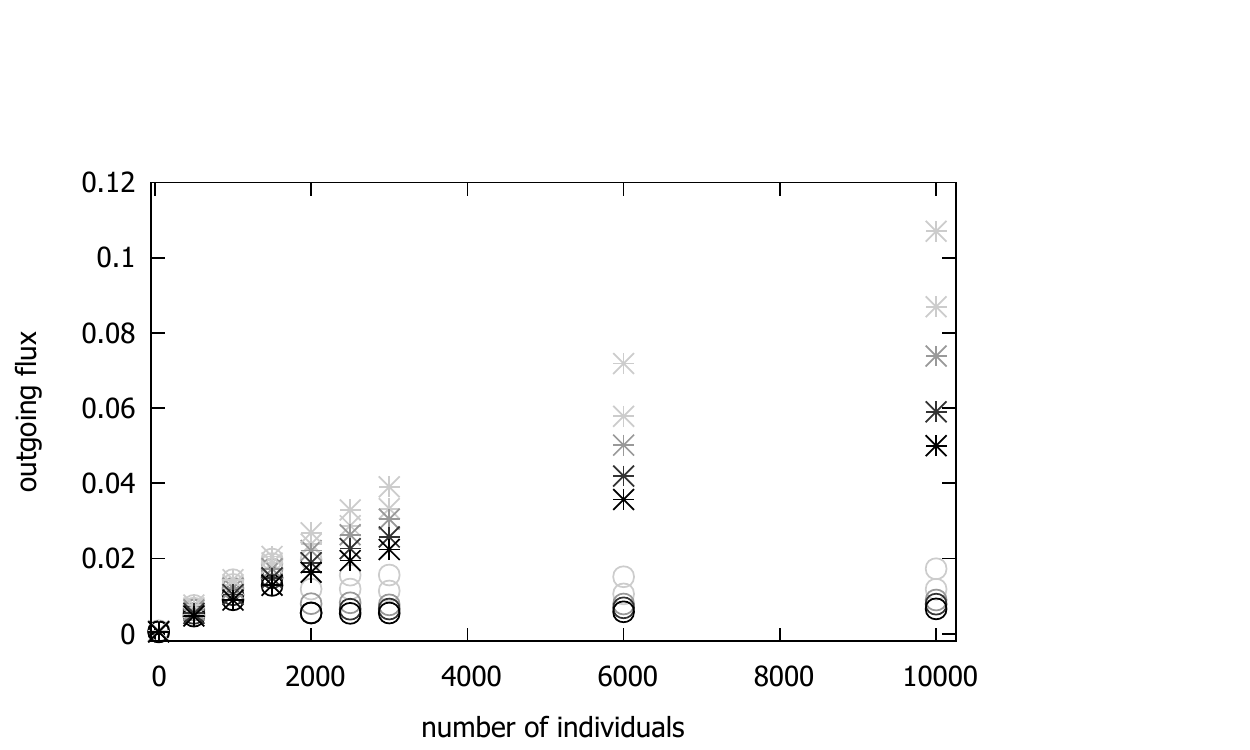}
}
\end{picture}
\caption{{\bf Averaged outgoing flux vs. number of pedestrians.}
Simulations with $L=101$, $R=1$, and, from the darkest to the lower grey, $W=0,\,1,\,3,\,5,\,10$.
On the left: plot of the average outgoing flux for $T=0$ ($+$).
On the right: plot of the average outgoing flux for $T=5$ ($\ast$) and $T=300$ (circles).
}\label{fig6}
\end{figure}

Our simulations show that the leading effect is the increase of the flux rate. In Fig.~\ref{fig6}, we show the effect of the wall interaction on the outgoing flux. We plot the measured average outgoing flux for different values of $W\geq 0$.  In the figure, to have a readable picture, we show the cases $T=0$ in the left panel, and the cases $T=5$ and $T=300$ in the right panel, but the increasing effect is present for any value of the threshold. We use grey tones to distinguish the values assumed by $W=0,1,3,5,10$, using lighter tone when $W$ is larger.
For any fixed  value of the threshold $T$, the flux increases monotonically 
with respect to $W$.
The increment is bigger when the clustering effect is not so strong, hence for small values of $T$. 

\begin{figure}[!ht]
\begin{picture}(400,140)(0,0)
\centering
\put(-6,0){
  \includegraphics[width=0.62\textwidth]{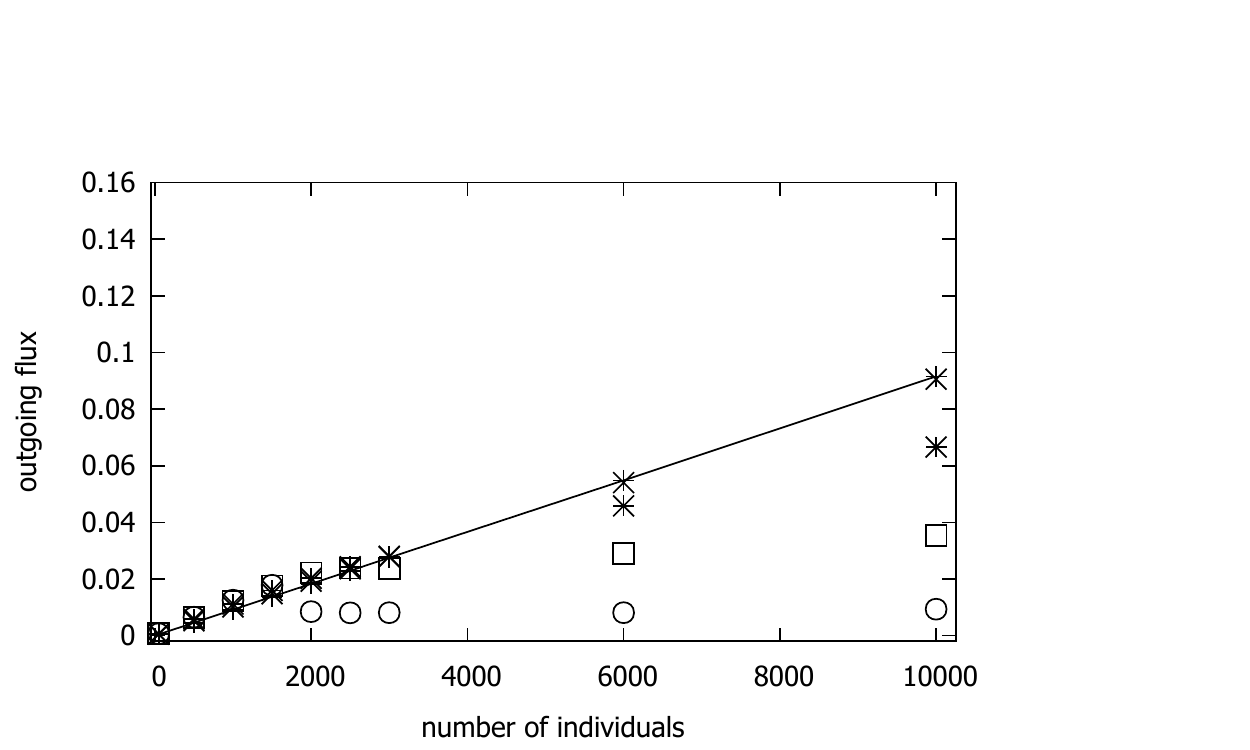}
}
\put(230,0){
  \includegraphics[width=.62\textwidth]{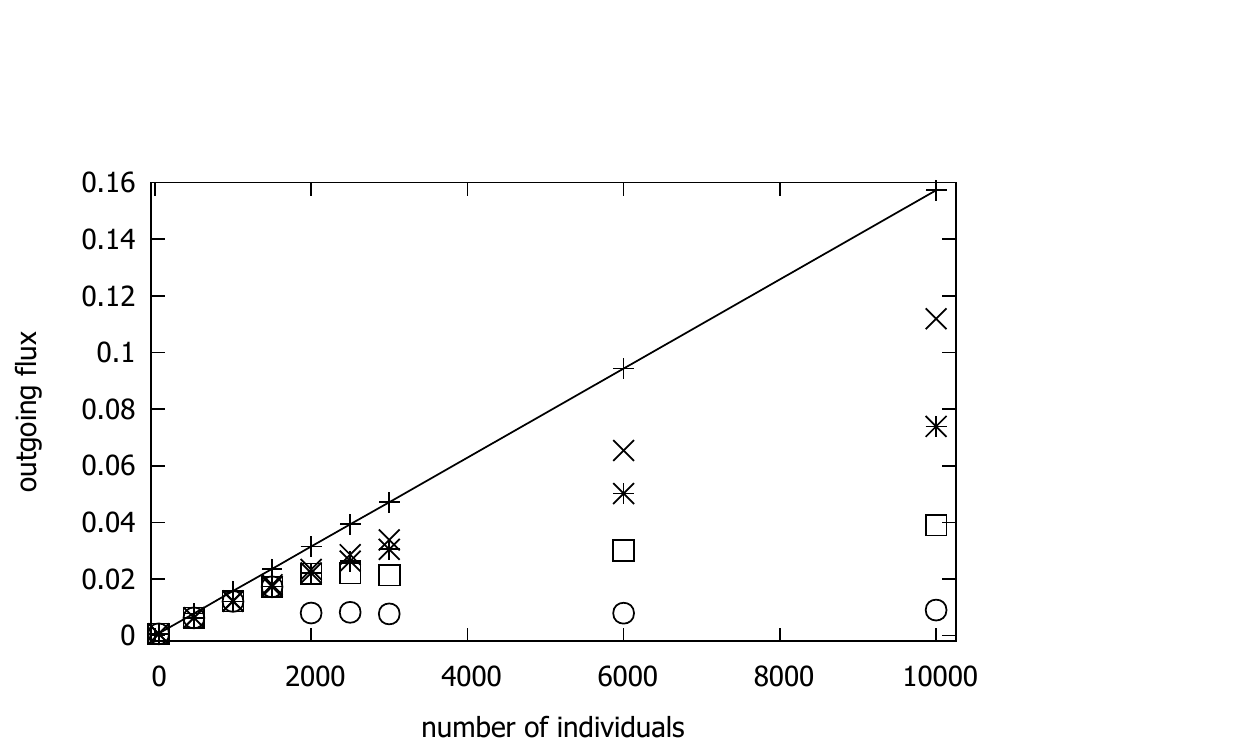}
}
\end{picture}
\caption{{\bf Averaged outgoing flux vs. number of pedestrians.}
The symbols $+$, $\times$, $\ast$, the squares and the circles represent respectively the cases $T=0,2,5,30,300$, with  $L=101$, with $W=3$, and $R=1$.
 On the left  we show the simulation results for the choice of the \emph{threshold exit}, on the right the data refer to the \emph{sure exit} rule.
}\label{fig7}
\end{figure}

It is interesting to observe that even if the measured value of the flux increases when $W$ grows, for any fixed value of $W$ the plot of the average flux as function of $N$ has the same behavior of the one resulting in the case $W=0$.
To illustrate this observation, we show  in Fig.~\ref{fig3n}  a plot analogous to the one we made for $W=0$.
The qualitative behavior of the outgoing flux as a function of $N$ does not change for $W>0$.
 We  see   in Fig.~\ref{fig7} illustrating the case $W=3$ that  with both the sure exit and threshold exit rules the plots are similar to those in 
Fig.~\ref{fig3n}.

We also notice  that even in this case, as observed earlier for $W=0$, the clustering regime starts for $T=300$ when the number of individuals grows up to $2000$. If a small number of particles ($N\leq1500$) is involved in the dynamics, then  the measured outgoing fluxes for any fixed value of the parameter $W$ are really close. As a consequence, the measured flux does not depend on $T$ for these values of $N$. 

We argued that the interaction with the wall is a plausible decision heuristic that could decrease the information overload associated with large population sizes. In line with this expectation, for low population size, the variation in interaction with the wall makes no substantial difference in the outgoing flux, supporting the observation that in simple environments, heuristics are not particularly useful. As information overload increases for population sizes larger than 2000, the small groups (T=5) seem to benefit more from the use of this decision heuristic than the large groups (T=300). However, as mentioned earlier, the outgoing flux seems to increase in all group sizes when interaction with the walls increase. So in general, this decision heuristic seems to be an effective way of coping with information overload, yet it cannot completely override the coordination costs associated with large groups.

\subsection{The role of the obstacle}
\label{s:l-obs}
\par\noindent
We consider now the presence of a square obstacle inside the corridor. The obstacle is expected to increase the coordination costs because it may wrongly signal an exit, and reduce freedom to move in the room. We vary two indices related to the obstacle, namely its size and its location as parameters that can represent the increase in coordination costs associated with the obstacle. When the object is small or located in the center its impact on coordination costs is expected to be lower than when the object is large and located close to the exit. The obstacle is represented by a number of sites in the corridor that are not accessible to the agents. Note that the sites in the corridor  neighboring the obstacle but accessible to the individuals will be consider as boundary sites, i.e., the agents might prefer to move close to this internal boundary due to the possible wall stickiness $W$.

We expect that  the obstacle can influence the resulting outgoing flux, because of excluded volume, in two different ways.
On the one hand  it can make easier or harder for a particle to find the exit depending on its position, for instance making more difficult to reach a region far from or close to the exit. On the other the presence of an obstacle can  favor the clustering formation and influence the region in which the cluster is formed.

\begin{figure}[!ht]
\begin{picture}(400,140)(0,0)
\centering
\put(-6,0){
  \includegraphics[width=0.62\textwidth]{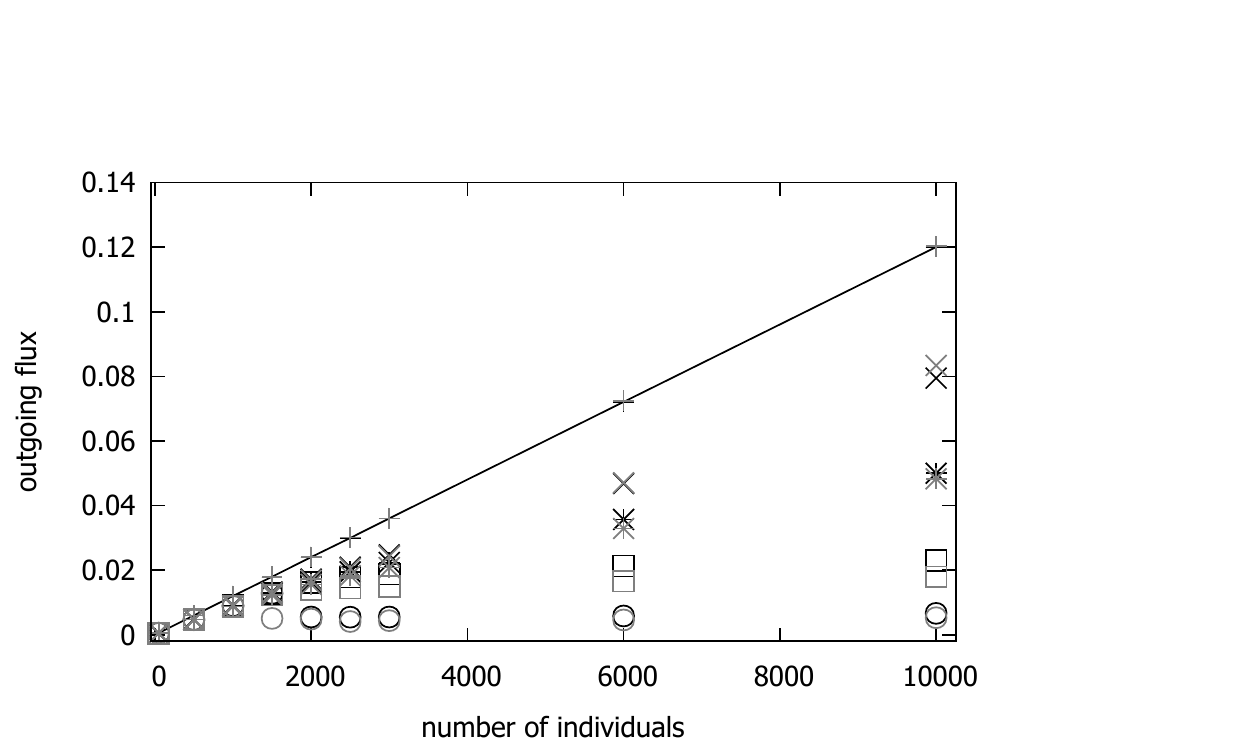}
}
\put(230,0){
  \includegraphics[width=.62\textwidth]{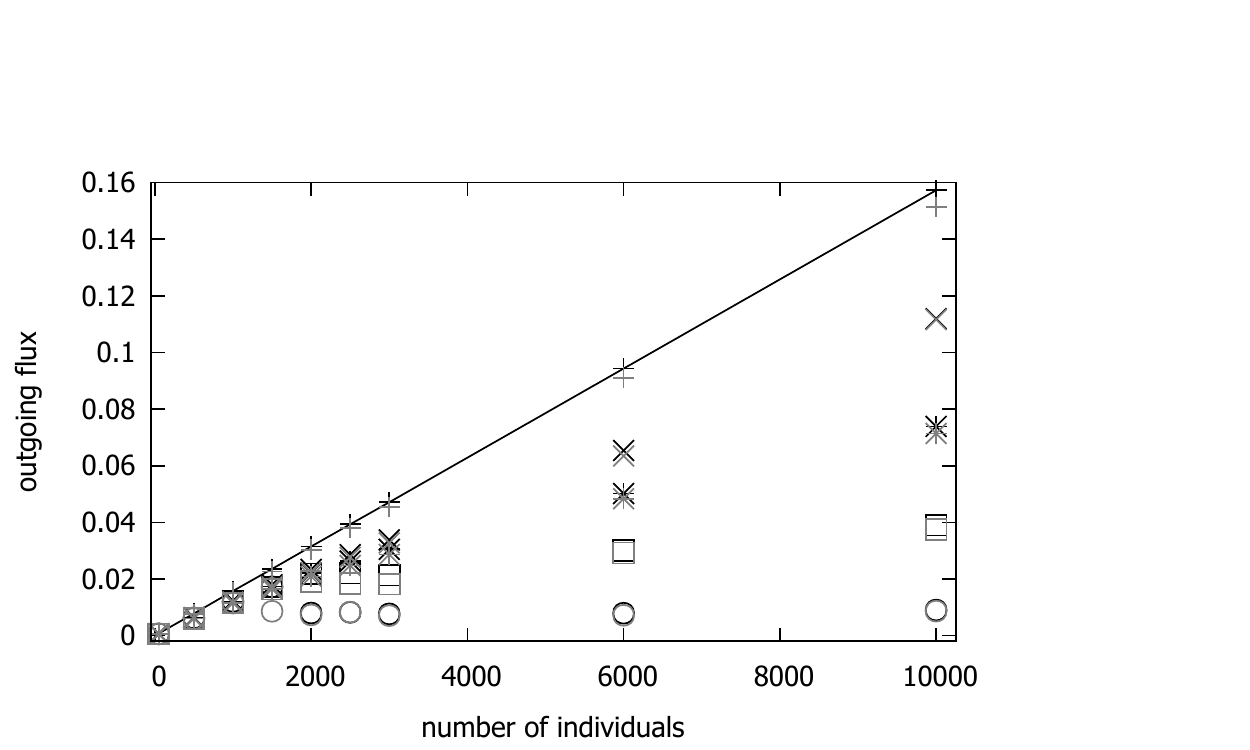}
}
\end{picture}
\caption{{\bf Averaged outgoing flux vs. number of pedestrians.}
The symbols $+$, $\times$, $\ast$, the squares and the circles represent respectively the cases $T=0,2,5,30,300$, with  $L=101$, $R=1$, \emph{sure exit} rule and a squared obstacle with side $41$ placed in the lattice with center in $(51,51)$.
The grey symbols represent the case of presence of the obstacle, while the black symbols refer to the empty lattice.
On the left  we show the simulation results for $W=0$,  on the right the data refer to  $W=3$.
}\label{fig_oc}
\end{figure}

We notice in our simulation that if the obstacle is sufficiently small it does not visibly affect  the flux, 
so we consider now the case of a squared obstacle with side $41$. 
 In Fig.~\ref{fig_oc} we show the behavior of the flux if it is  present this squared obstacle with side $41$ with center in the center of the corridor.

Small values of the threshold lead to small variations in the average exit flux. We can see this trend for $T=0$, $2$, and $5$.
It is instead interesting to note the clustering induced by the obstacle for large values of $T$. The flux drop for $T=300$ happens for $N=1500$ both if $W=0$ and if $W=3$. We recall that, in the absence of the obstacle, we observe this flux drop for $T=300$ for the first time around $N=2000$ (see Fig.~\ref{fig1n}--\ref{fig2n}).
For $T=30$ the flux in the case $W=0$ is reduced by the cluster formation (see the squares in Fig.~\ref{fig_oc}, in the left panel, the grey symbols are always below the black).

\begin{figure}[!ht]
\begin{picture}(400,140)(0,0)
\centering
\put(-6,0){
  \includegraphics[width=0.62\textwidth]{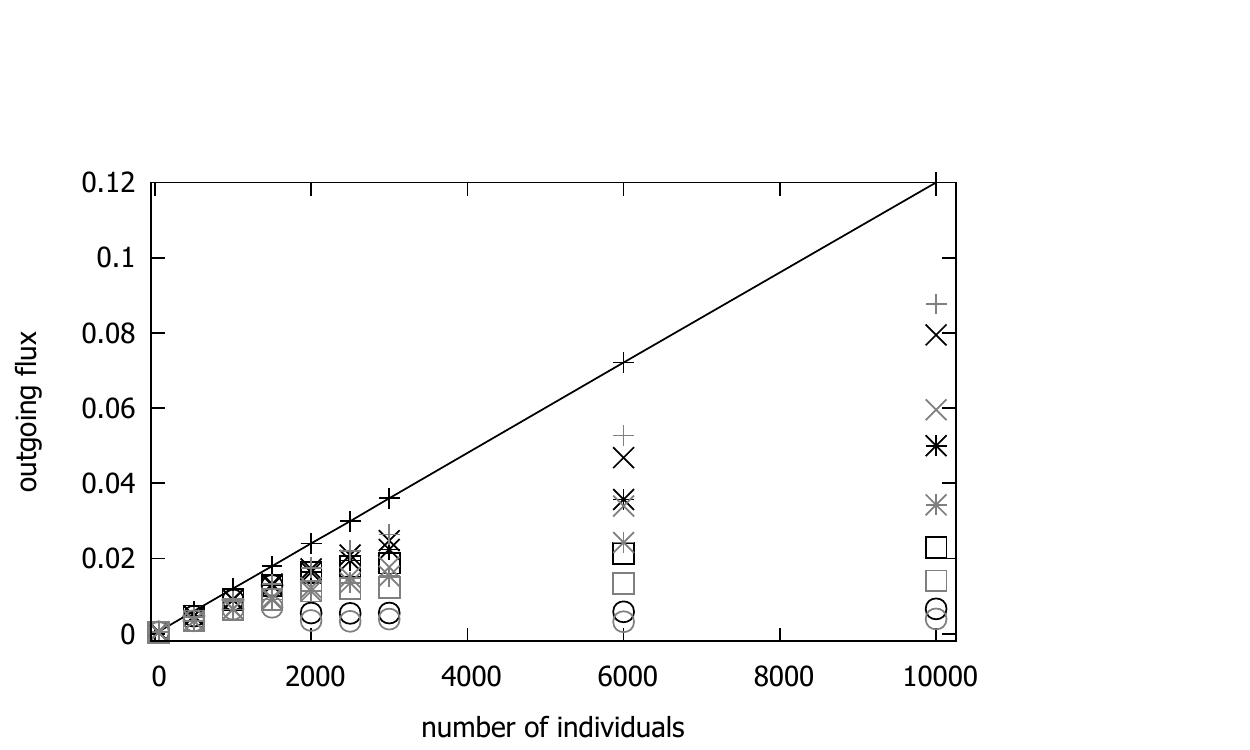}
}
\put(230,0){
  \includegraphics[width=.62\textwidth]{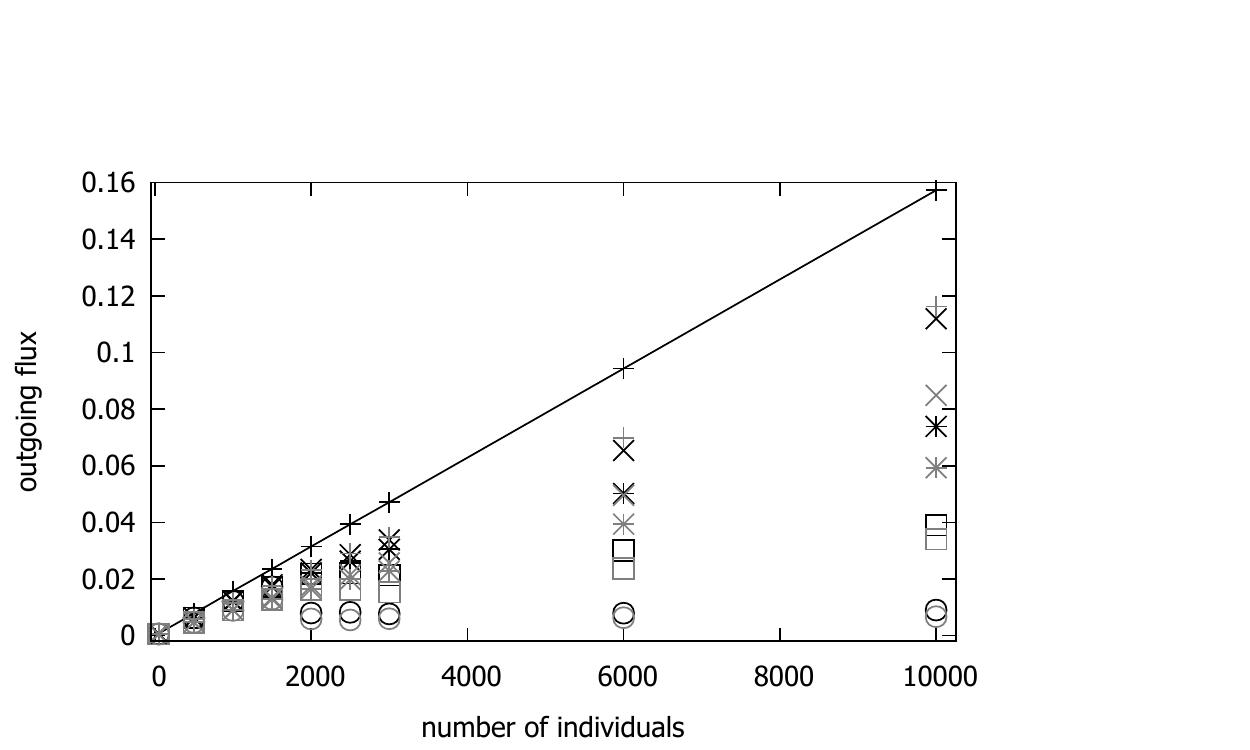}
}
\end{picture}
\caption{{\bf Averaged outgoing flux vs. number of pedestrians.}
As in figure \ref{fig_oc}, with a squared obstacle with side $41$ and center in $(71,51)$.
}\label{fig_or}
\end{figure}

We find  a strong asymmetry in the positioning of the obstacle with respect to the center of the corridor.
Let us first consider the same $41\times41$  obstacle, but we position it to be closer to the exit than to the opposite side of the lattice.
 In Fig.~\ref{fig_or}, we show the data for the square  with center in $(71,51)$. 
 We find out that this obstacle close to the exit results in  a loss in the outgoing flux (in figure we find the grey symbols always below the blacks, both if $W=0$ and  $W=3$).
 It is interesting to notice that  now the clustering effect does not appear for smaller $N$ compared to the case of the  corridor without obstacle. 

\begin{figure}[!ht]
\begin{picture}(400,140)(0,0)
\centering
\put(-6,0){
  \includegraphics[width=0.62\textwidth]{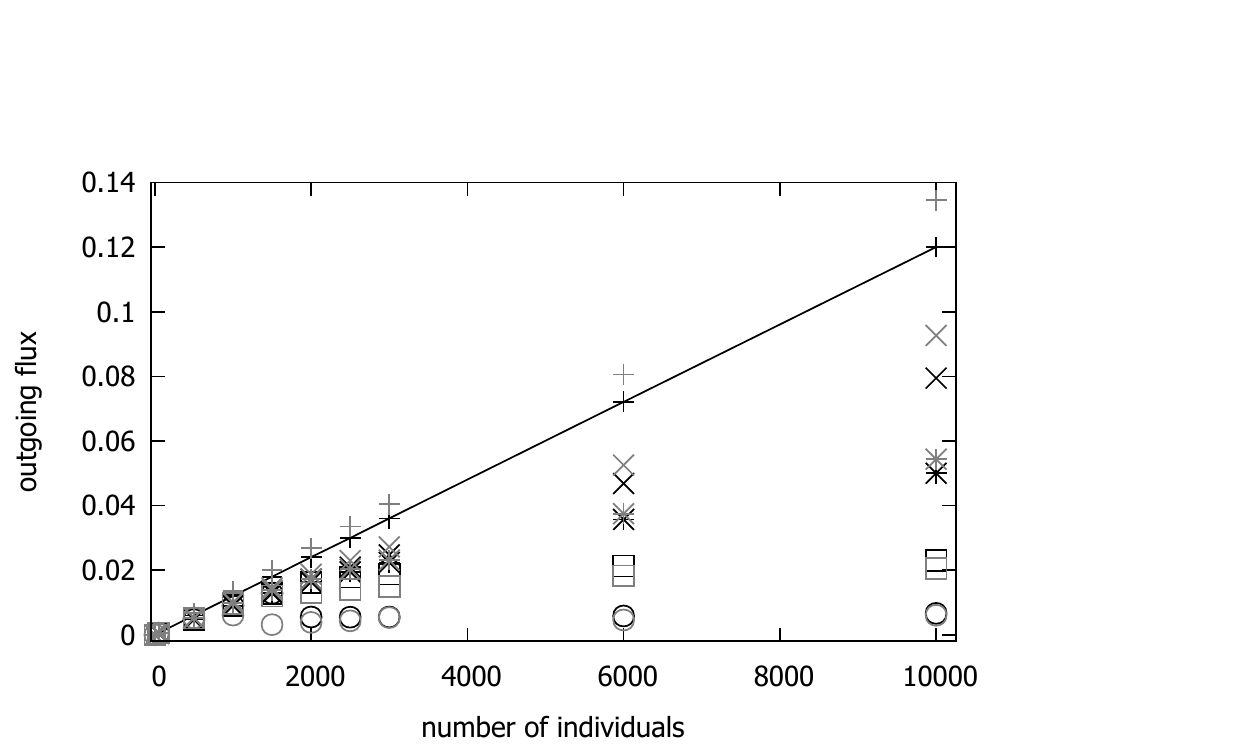}
}
\put(230,0){
  \includegraphics[width=.62\textwidth]{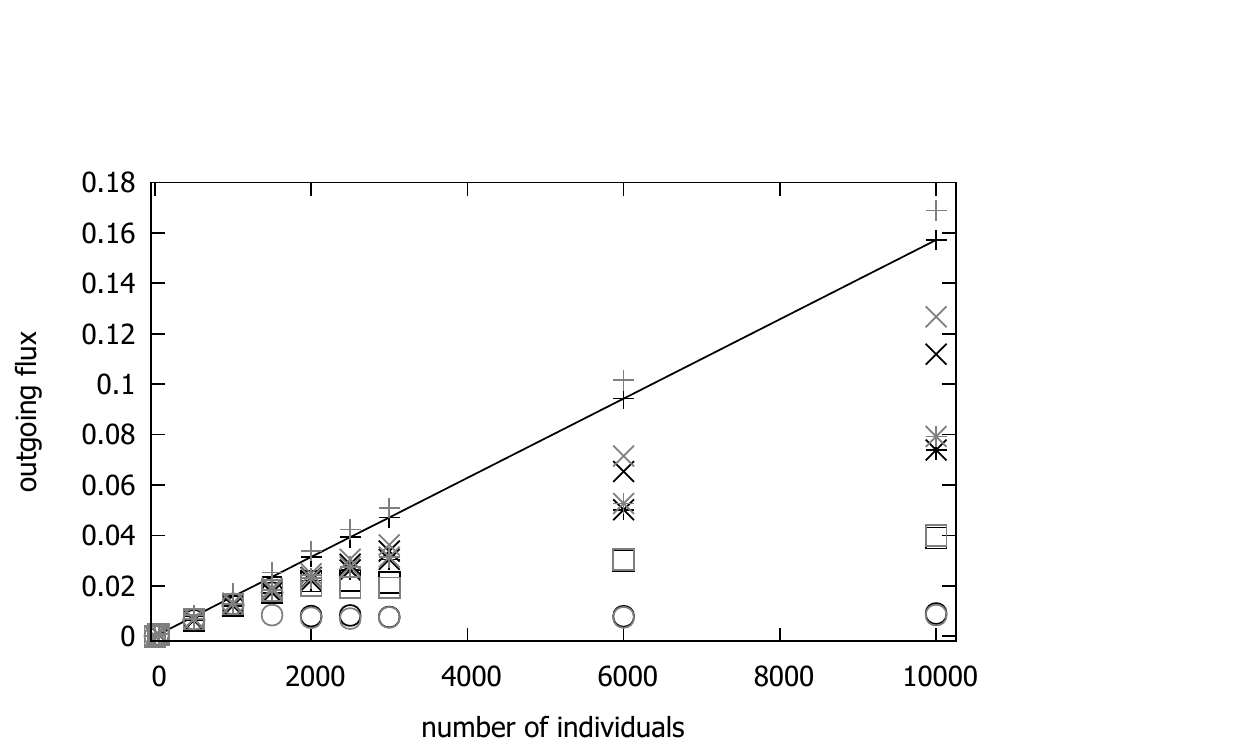}
}
\end{picture}
\caption{{\bf Averaged outgoing flux vs. number of pedestrians.}
As in Fig.~\ref{fig_oc}, with a squared obstacle with side $41$ and center in $(31,51)$.
}\label{fig_os}
\end{figure}

As expected, we observed in this case that an obstacle close to the exit cannot increase the average flux.
Consider now the square with center in $(31,51)$.
We see in Fig.~\ref{fig_os} that, for this particular choice of geometry,  the outgoing flux increases  if the threshold is  not too large ($T\leq 5$ in the plot).
On the contrary, if the threshold is large, $T\geq 30$ in the figure, we can observe that the clustering regime is reached earlier than in the empty corridor case.
Indeed, there is a loss in the outgoing flux for $T=30$ and $W=0$, and the clustering appears at $N=1000$ for the parameters $T=300$ and $W=0$, while at $N=1500$ for $T=300$ and $W=3$. 

Summarizing, we find a strong dependence on the position of the obstacle. If the obstacle is closer to the exit than to the opposite side, the flux decreases.
On the contrary, if the obstacle is close to the entrance and far from the exit, the flux increases for  small $T$. 
 Interestingly, if the threshold is large,  an obstacle close to the entrance favors the formation of clusters.

Our interpretation of the results is the following: The obstacle, placed close to the exit, makes more difficult to reach the region on its right side where there the exit location is, and, it produces a decrease in the outgoing flux with respect to the empty lattice case.

An obstacle close to the other side makes difficult for a particle to reach again the region on the left side of the obstacle once it is on the right side. It keeps the particles to spend longer time in the region with the exit, favorizing the exit.
However, since the obstacle reduces the free region close to the entrance point, it makes more likely the formation of a big stable cluster in that region. This may produce a decrease in the flux, when coordination constraints are significant, namely if the group size is sufficiently big and the population size is sufficiently large. In other words we find that an obstacle adds coordination constraints when is large in size, situated next to the exit and the population and group sizes are large. We find that the outgoing flux is impaired in overcrowded spaces, with large clusters formed and in which a large obstacle is present next to the exit. Efficient information integration in small--sized groups can overcome the coordination constraints added by an obstacle, especially if this is not located next to the exit. This novel insights lends support for our initial interpretation that smaller groups are more effective information processing units and cope easier with the coordination problems.


\subsection{Evacuation time}
\label{s:l-eva}
\par\noindent
We are interested  in evaluating  the evacuation time, i.e., the average time needed to let all the individuals leave the corridor.

We  consider  the following experiment:
We dispose $N$ pedestrians randomly in the corridor and let the dynamics start accordingly with the description of the model in Sect. \ref{s:l-model}, but we do not consider any new individual entering in the corridor. We observe the time needed until the last pedestrian will exit the corridor.
The average time measured repeating this experiment will be called \emph{evacuation time}.

The results discussed concerning the outgoing flux allow us  to argue which behaviors would be observed for the evacuation time. Note that since the number of agents is decreasing during the evolution of the system, it will be more difficult to enter in a clustering regime (i.e., it is necessary a larger number of particles to reach it) and the coordination constraints in the system are likely to be lower especially for small group and population sizes. Building on the results on the effect of an obstacle on the outgoing flux, we expect to find analogously that the evacuation time is smaller when the corridor is empty and if an obstacle is far from the exit. On the contrary, the closer an obstacle is to the exit, the longer it will take for all particles to exit the room.

\begin{figure}[!ht]
\begin{picture}(400,140)(0,0)
\centering
\put(-6,0){
  \includegraphics[width=0.62\textwidth]{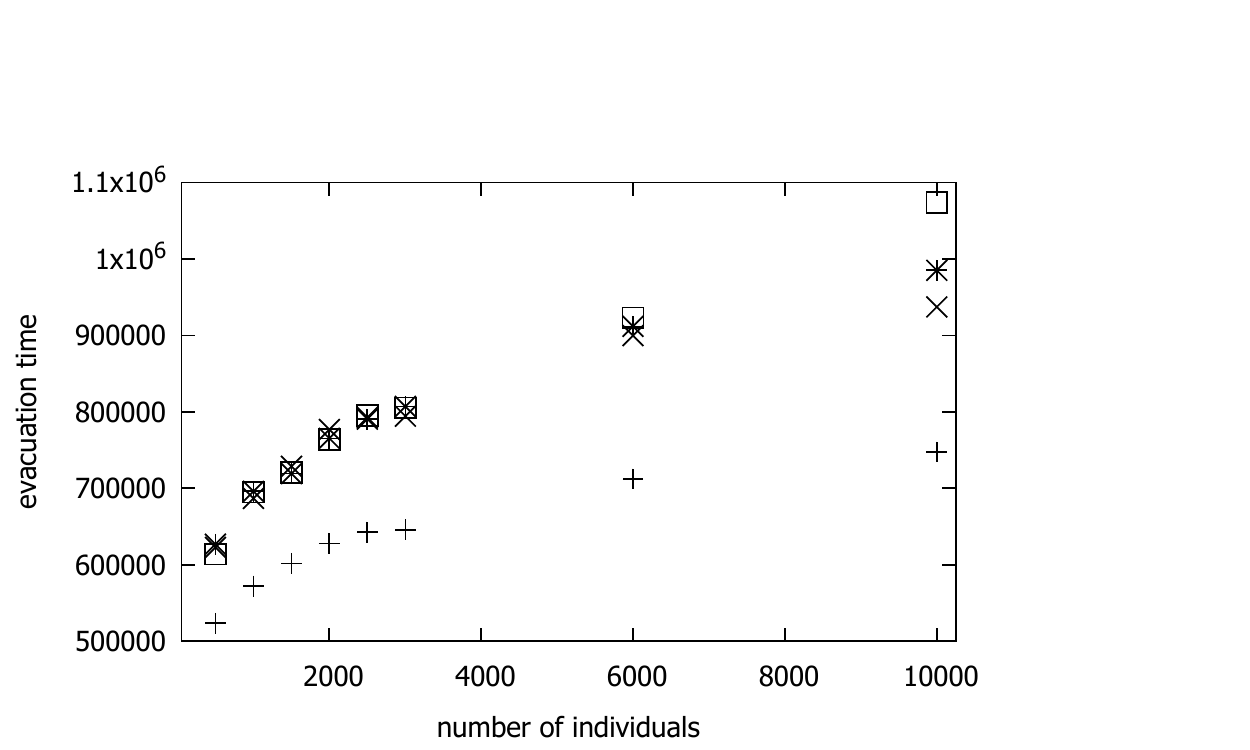}
}
\put(230,0){
  \includegraphics[width=.62\textwidth]{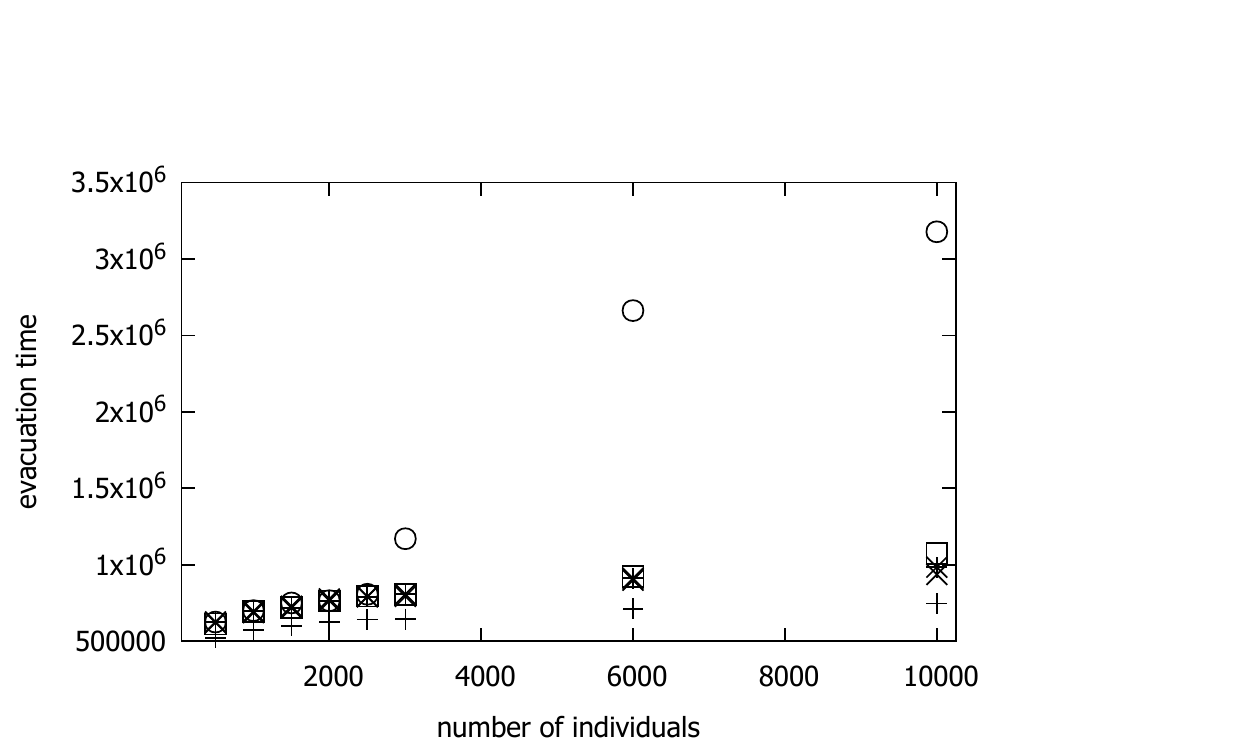}
}
\end{picture}
\caption{{\bf Averaged evacuation time vs. number of pedestrians.}
The symbols $+$, $\times$, $\ast$, the squares and the circles represent respectively the cases $T=0,2,5,30,300$, with  $L=101$, $W=0$, $R=1$, sure exit and no obstacles in the corridor.
 On the left we excluded the results for $T=300$ to have a more readable figure, on the right we have the same picture added of the data for $T=300$.
}\label{fig_ev1}
\end{figure}

We plot in Fig.~\ref{fig_ev1} the evacuation time as a function of the number of individuals $N$, for $W=0$, and values for $T$ from $0$ to $300$. The geometry is the corridor without the obstacle.

We observe that the evacuation time is monotone with respect to $T$ for a fixed $N$ if $N$ is large, while is essentially not depending on $T>0$ if $N$ is small. The independent random walk for $T=0$ guarantees the minimum evacuation time.
 Note also that the system reaches the clustering regime only for large values of the threshold $T$ and for large $N$, for
 $N\geq 3000$ when $T=300$ in the plot. 
Note that in this clustering regime there is a clear growth in the evacuation time.

\begin{figure}[!ht]
\begin{picture}(400,140)(0,0)
\centering
\put(-6,0){
  \includegraphics[width=0.62\textwidth]{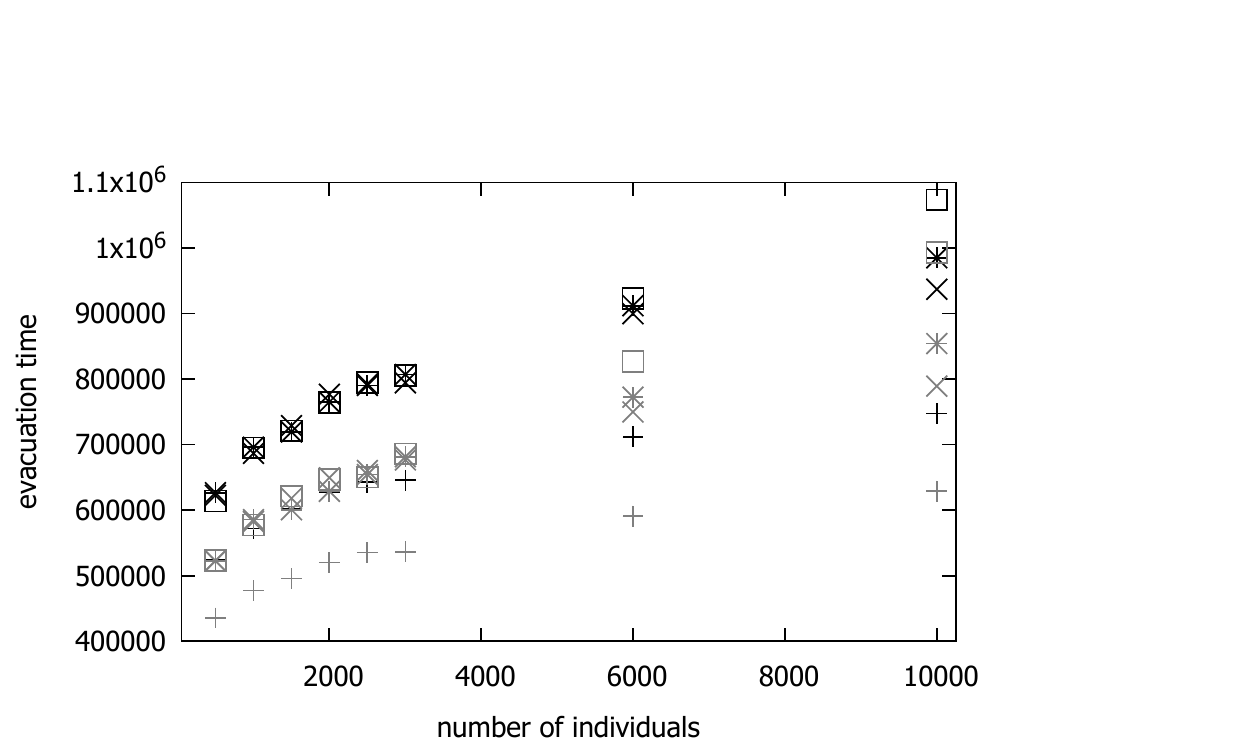}
}
\put(230,0){
  \includegraphics[width=.62\textwidth]{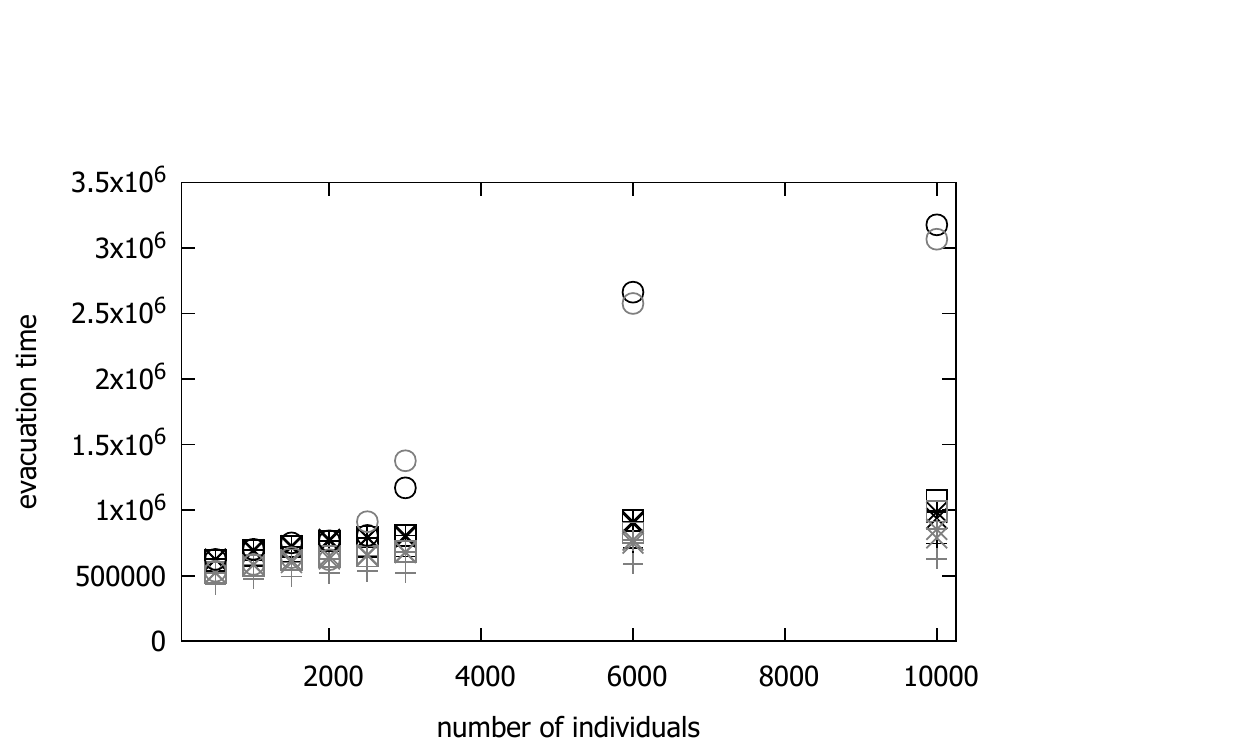}
}
\end{picture}
\caption{{\bf Averaged evacuation time vs. number of pedestrians.}
The symbols $+$, $\times$, $\ast$, the squares and the circles represent respectively the cases $T=0,2,5,30,300$, with  $L=101$, $W=0$, $R=1$, sure exit and a squared obstacle with side $41$ and center in $(31,51)$ in the corridor.
The gray symbols represent the case of presence of the obstacle, while the black symbols refer to the empty lattice.
 On the left we excluded the results for $T=300$ to have a more readable figure, on the right we have the same picture added of the data for $T=300$.
}\label{fig_ev2}
\end{figure}

\begin{figure}[!ht]
\begin{picture}(400,140)(0,0)
\centering
\put(-6,0){
  \includegraphics[width=0.62\textwidth]{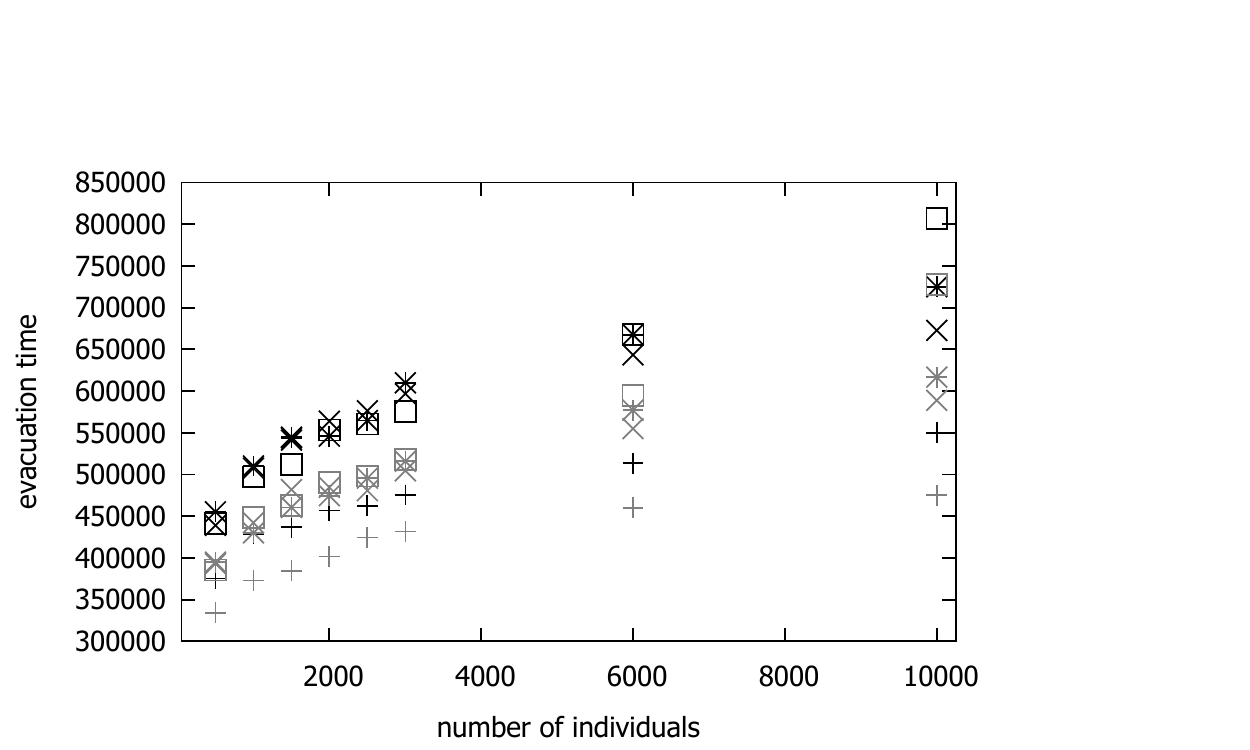}
}
\put(230,0){
  \includegraphics[width=.62\textwidth]{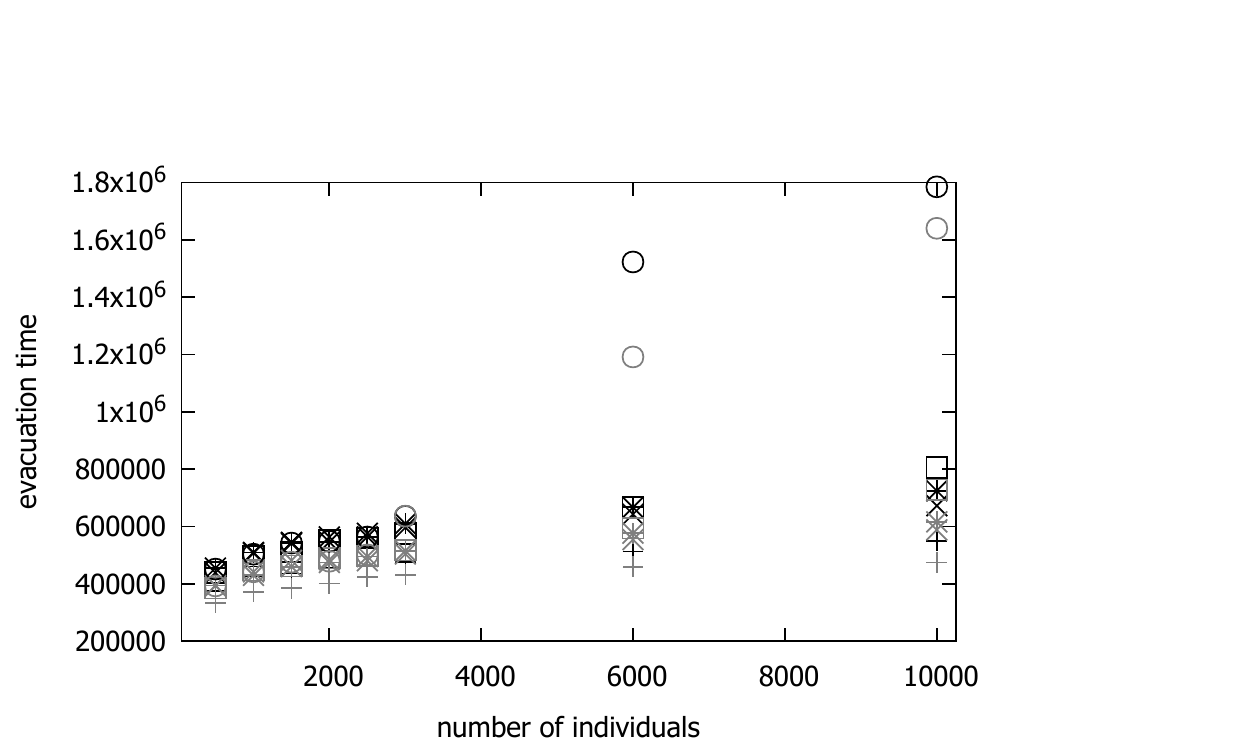}
}
\end{picture}
\caption{{\bf Averaged evacuation time vs. number of pedestrians.}
As in Fig.~\ref{fig_ev2} with $W=3$.
}\label{fig_ev3}
\end{figure}

\begin{figure}[!ht]
\begin{picture}(400,140)(0,0)
\centering
\put(-6,0){
  \includegraphics[width=0.62\textwidth]{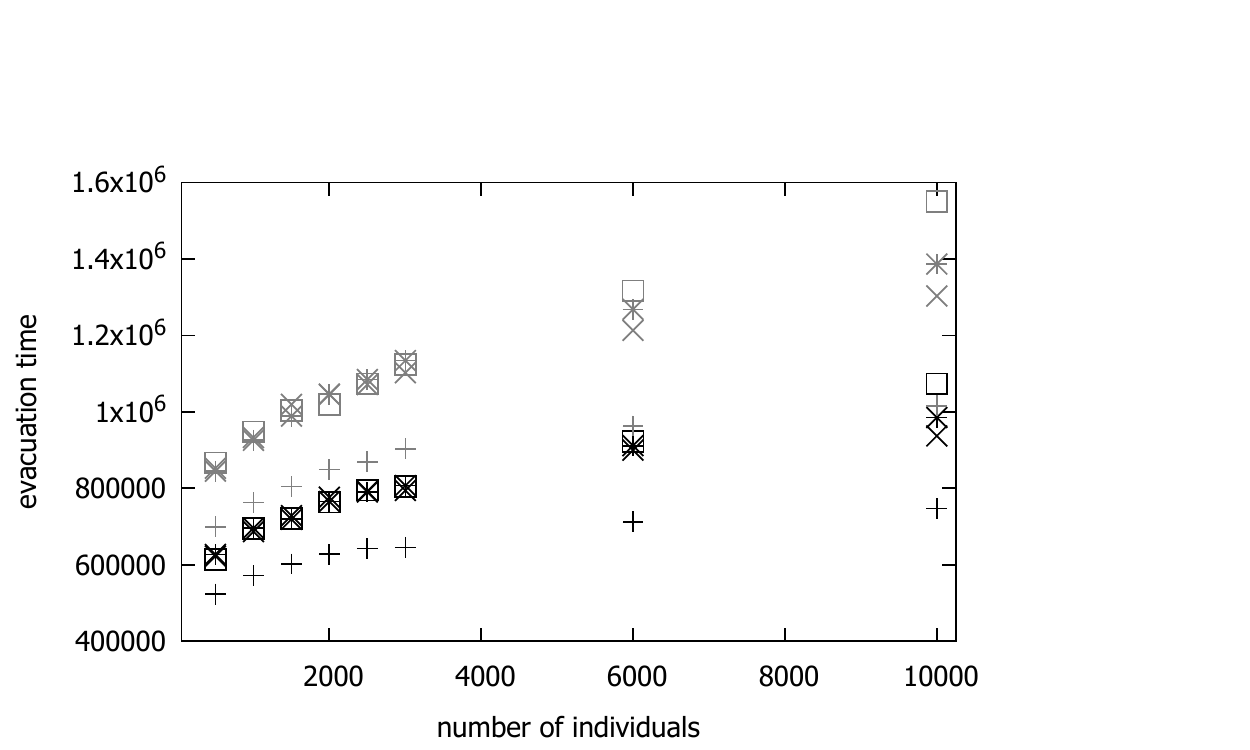}
}
\put(230,0){
  \includegraphics[width=.62\textwidth]{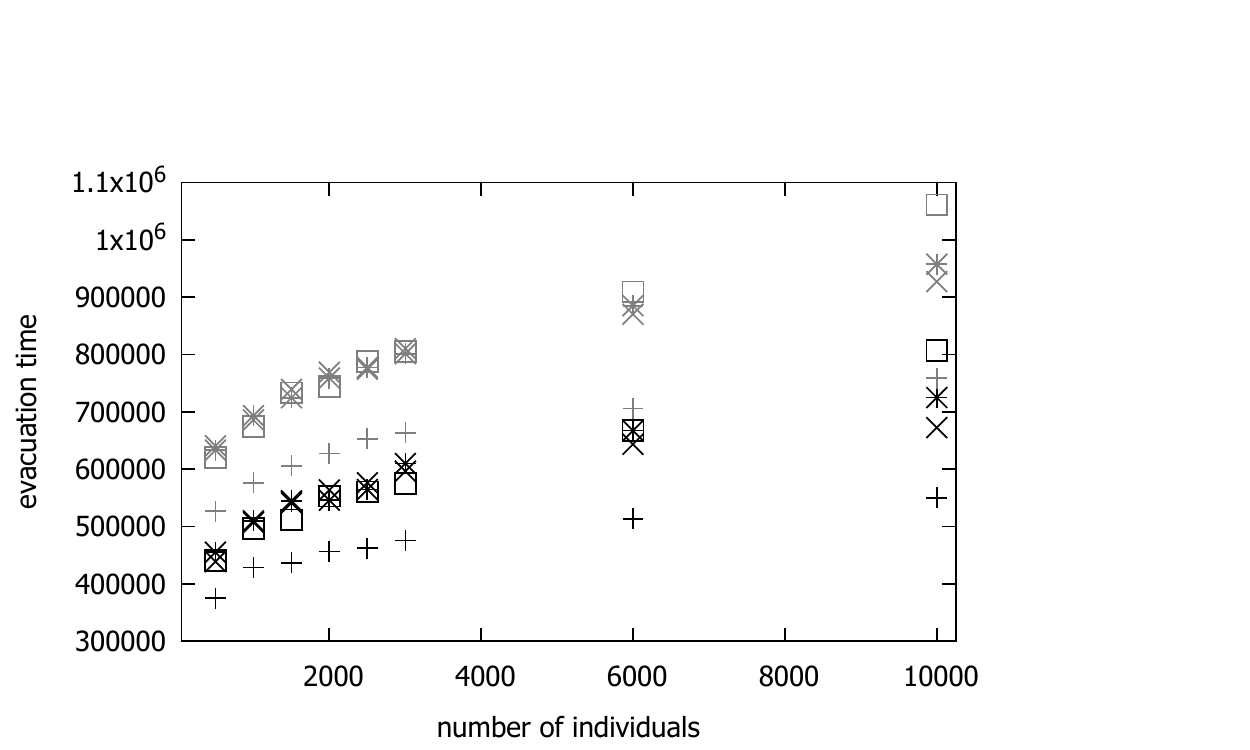}
}
\end{picture}
\caption{{\bf Averaged evacuation time vs. number of pedestrians.}
The symbols $+$, $\times$, $\ast$ and the squares  represent respectively 
the cases $T=0,2,5,30$, with  $L=101$, $R=1$, sure exit and a squared obstacle with side $41$ 
and center in $(71,51)$ in the corridor.
The gray symbols represent the case of presence of the obstacle, while the 
black symbols refer to the empty lattice.
 On the left  we show the simulation results for $W=0$,  on the right the data 
refer to  $W=3$.
}\label{fig_ev4}
\end{figure}

The presence of the obstacle leads to expected effects on the evacuation time.
As discussed in the previous section, we consider a large squared obstacle with side $41$ and we position it to be either in the left half of the corridor or in the right half.
We report in Fig.~\ref{fig_ev2}--\ref{fig_ev3} the average evacuation time calculated for the empty corridor (black symbols) and for the presence of the obstacle of side $41$ with center in $(31,51)$ as in Fig.~\ref{fig_os} (grey symbols).
We consider the cases of no influence by the wall on the dynamic ($W=0$)  in 
Fig.~\ref{fig_ev2} and the case of positive $W$ in Fig.~\ref{fig_ev3}. The presence of the obstacle in the left half of the corridor, far from the exit, produces a decrease in the evacuation time in most of the considered data. It is interesting to note that for $W=0$ and $T=300$ (Fig.~\ref{fig_ev2}, right picture) the clustering effect starts for a smaller number of agents with respect to the empty corridor case, so for $T=300$ and $N$ between $2500$ and $3000$ the evacuation time is longer if the obstacle is present. Once the clustering regime is reached in both cases (large $N$), the obstacle yields again a shorter evacuation time.

In Fig.~\ref{fig_ev4} we consider the presence of the same square obstacle in the right half of the corridor, closer to the exit. We compare the evacuation time in presence of this obstacle (gray symbols) and in absence of obstacles (black symbols).
The center of the obstacle is now put in $(71,51)$. As in the case of the evaluation of the outgoing flux we find this positioning to be unfavorable, i.e., it produces an increment in the evacuation time, both if $W=0$ (left picture) and if $W=3$ (right picture).

\subsection{Discussion}
\label{s:l-dis}
\par\noindent
As population size increases the likelihood of information complexity in the general system increases as well (a large number of individuals develop partial representations about an otherwise ambiguous environment), therefore the benefit of higher information pool in subgroups decreases in importance. In other words, groups of minimum size take advantage of low coordination costs and information redundancy in the general population in order to speed up evacuation. Therefore, for large population size, the information complexity of the whole system overrules the benefit of information sharing in small groups varying in size. These results are somehow similar with the simulation results reported 
by \cite{couzin} 
showing that as the population size increases the number of informed individual needed to direct the flock decreases. 
For R=0 individuals are required to move while for R=1 the probability of individual movement imposed is lower. This constraint increases the likelihood of exit especially for subgroups of smaller size (dyads and groups of five). This result supports the argument that informational complexity available in the system propagates better if individuals are asked to change positions as compared with the situation in which they can preserve their position on the lattice. Therefore, for systems of high information complexity (i.e., large population size) the probability of movement increases the rate of exit for smaller groups more than for large groups. As individuals (are forced to) move they acquire information from others in the room and, consequently, they will develop faster accurate interpretations concerning the availability of exit routes. To conclude, increasing the probability that individuals will freely move in the room, increases the rate of exit, by fostering information exchange that will ultimately accentuate the overriding effect of whole system’s information complexity over the small group information pool.

As we argued before, the informational complexity of the whole system generates a blockage for large groups. Large groups seem to be effective in integrating informational inputs in simpler rather than highly complex systems 
as indicated in 
Fig.~\ref{fig1n} 
where the exit rate increases with $T$ 
for population size under $1500$. 
Under situations of informational scarcity, 
group size seems to be a prerequisite for 
effective evacuation simply because the 
population size does not allow the formation 
of very large groups. In highly complex systems, 
the coordination costs (it becomes too difficult 
to find ways in which individual actions are to be 
synchronized in the whole system) and information 
processing costs (there are too many different 
information inputs that need to be integrated and 
the cluster is basically clogged) become unsurmountable 
for extremely large groups ($T=300$).  

Dyads, requiring minimal coordination costs, 
are better able to integrate effectively the 
(otherwise diverse) individual information inputs 
generated by frequent interactions especially imposed under $R=0$. 
The monotonic flattening of the exit rate graph for $T=30,5,2$ 
(rather than a sudden decrease as for $T=300$) 
shows that these small group configurations are actually 
the only viable (social) information processing units in 
highly complex social environments. If large groups are 
more effective information integrators in rather simple 
social environments marked by ambiguity, dyads become 
the most effective information integrators in ambiguous 
environments with high social complexity.   

The results for the interaction with the walls resemble following some sort of decision heuristic (it is a higher chance of finding an exit in a wall than in other locations in the room) so following this heuristic circumvents the added value of larger groups sharing information in a particular location. 

The results show that effective information processing in small groups overrides coordination costs only in small populations. Moreover, the use of decision heuristics (i.e., presence of a wall) seems to increase the efficient information processing especially in small groups formed in dense population spaces. Coordination costs seem to be clearly disadvantageous especially when the group size is very large. In other words information integration in very large groups will not override the coordination costs associated with group size. Moreover, additional coordination constraints, like obstacles situated next to the exit, will further increase the negative effects of coordination costs both on outgoing flux as well as the evacuation time.

\section{Conclusions}
\label{s:conclusions}
\par\noindent
This study started from the observation that in ambiguous situations people use
others to validate the information they have and this process plays an
important role in situations in which a crowd needs to evacuate a darkened
space. We claimed that the information pool (how many members are exposed to
stimuli) and group’s computational resources (how many members actually
evaluate informational inputs) are two key factors influencing the speed of
evacuation. 

Our simulation results suggest that few information sources require
large clusters, in other words, scarcity of information requires large
computational resources. Moreover, we show that many information sources
require small groups for effective evacuations as information redundancy
requires little computational resources. As information resources become
redundant, due to the high coordination costs in large groups and social
loafing processes large groups become actually clogged by redundant information
and evacuation rate decreases substantially.

A direct policy implication of our results is to stimulate the formation of very small group sizes (up to 5 members) when people are trapped in confined and darkened spaces, as this seems to foster both the outgoing flux as well as the evacuation time for the whole population.

\begin{acknowledgments}
ENMC and AM thank ICMS (TU/e, Eindhoven, The Netherlands) for the 
kind hospitality and for financial support. PLC was supported by a grant of the Romanian National Authority for Scientific Research, CNCS -- UEFISCDI, project number PN--III--P4--ID--ERC--2016--0008.
The funders had no role in study design, data
collection and analysis, decision to publish, or 
preparation of the manuscript.
\end{acknowledgments}

\end{document}